\documentclass[aps, nofootinbib]{revtex4}
\usepackage{graphicx}

\newcommand{\Ref}[1]{(\ref{#1})}

\def \f{\frac}

\def \be{\begin{equation}}
\def \ee{\end{equation}}
\def \bes{\begin{eqnarray}}
\def \ees{\end{eqnarray}}
\def \arr{\rightarrow}

\newcommand{\Rb}{{\rm \bf R}}
\newcommand{\Cb}{{\rm \bf C}}

\newcommand{\Z}{{\rm \bf Z}}
\def \w{\wedge}
\newcommand{\Cm}[3]{C^{j_{#1}j_{#2}j_{#3}}_{m_{#1}m_{#2}m_{#3}}}
\newcommand{\C}[4]{C^{j_{#1}j_{#2}j_{#3}}_{#4_{#1}#4_{#2}#4_{#3}}}
\newcommand{\Cq}[6]{C^{J_{#1}J_{#2}J_{#3}J_{#4}#6}_{#5_{#1}#5_{#2}#5_{#3}#5_{#4}}}
\def \sl{SL(2,\Cb)}
\def \SA{{\cal A}}

\begin{document}

\title{Implementing causality in the spin foam quantum geometry}
\author{{\bf Etera R Livine}\thanks{e-mail: livine@cpt.univ-mrs.fr} \\
{\small Centre de Physique Th{\'e}orique,Campus de Luminy, Case 907,} \\
{\small 13288 Marseille cedex 9, France; \\ e-mail: livine@cpt.univ-mrs.fr}
\vspace{4mm} \\
{\bf Daniele Oriti}\thanks{e-mail: d.oriti@damtp.cam.ac.uk} \\
{\small Department of Applied Mathematics and Theoretical Physics,} \\
{\small Centre for Mathematical Sciences, University of Cambridge,} \\
{\small Wilberforce Road, Cambridge CB3 0WA, UK; \\ e-mail: d.oriti@damtp.cam.ac.uk}}

\begin{abstract}
We analyse the classical and quantum geometry of the Barrett-Crane
spin foam model for  four dimensional quantum gravity, explaining why
it has to be considering as a covariant realization of the
projector operator onto physical quantum gravity states. We
discuss how causality requirements can be consistently implemented
in this framework, and construct causal transiton amplitudes
between quantum gravity states, i.e. realising in the spin foam
context the Feynman propagator between states. The resulting
causal spin foam model can be seen as a path integral quantization
of Lorentzian first order Regge calculus, and represents a link
between several approaches to quantum gravity as canonical loop
quantum gravity, sum-over-histories formulations, dynamical
triangulations and causal sets. In particular, we show how the
resulting model can be rephrased within the framework of quantum
causal sets (or histories).

\end{abstract}
\maketitle

\newpage
\tableofcontents

\newpage

\section{Introduction}
Spin foam models \cite{Baez}\cite{danrev} have emerged recently as a new promising approach to the
construction of a quantum theory of gravity, and
much work has been devoted to the development of such models. However,
much remains to be understood: in particular, the way
these models actually encode the (quantum) geometry of spacetime is only
partially under control, and also the role of causality (that
we argue is indeed fundamental) in the existing models has been
investigated only to a limited extent. In this paper we
tackle both these issues, concentrating on the Lorentzian Barrett-Crane
model for 4-dimensional quantum gravity \cite{bc1}\cite{bc2}, after a
preliminary discussion of the 3-dimensional Ponzano-Regge model as an
illustrative example, and we study its quantum geometry, its symmetries
and the classical description of spacetime geometry
it corresponds to; then we discuss its causal properties, and show how it
can consistently be modified to implement causality
requirements in full and to define causal transition amplitudes between
quantum states of geometry. In this way it can be
shown to fit within the general scheme of causal spin foam models proposed
by many authors, being its first (highly) non-trivial example.

\
\
\subsection{Long-term plan and motivation}
Let us first give a few motivations both for spin foam models in general
and for our work in particular, and also outline a
long-term plan in which our results may be seen as a step forward.

First of all, there are several reasons to believe that an
approach which mantains full covariance in treating the
gravitational field is to be preferred over any other which breaks
this covariance \cite{SorkFork}, e.g. a canonical approach based
on a 3+1 splitting of the spacetime coordinates, or restricted to
particular topologies such as the customary $\Sigma\times\Rb$.
Even looking at the problem from the canonical side, the problems
encountered in implementing the Hamiltonian constraint in loop
quantum gravity \cite{carlorev}\cite{thi}, i.e. in understanding
the dynamics of gravitational states, and the possibility of
implementing it in a covariant manner by a projector operator
encourage the search for a \lq\lq covariant" version of loop
gravity, which is a way to look at spin foam models
\cite{mike&carlo}\cite{carloproj}. These can be seen as a peculiar
implementation of the path integral approach to quantum gravity,
where a partition function is defined as a sum over all the
4-geometries interpolating between given boundary 3-geometries,
with a weight given by the exponential of ($i$ times) the
Einstein-Hilbert action for general relativity, and a suitable
measure on the space of 4-metrics up to diffeomorphisms, with a
possible additional sum over all the possible manifolds having the
given boundary: \be
Z(h_1,h_2)\,=\,\sum_\mathcal{M}\int_{h_1,h_2}\,\mathcal{D}g\,e^{i\,S_{gr}}.
\ee Transition amplitudes between quantum states, as well as
expectation values of operators representing physical observables,
are computed by means of this partition function. Now the path
integral approach faces several problems, so that it is not even
clear how to make sense of the formal expressions it involves. One
may argue that these problems are the result of the use of
continuum geometric structures such as the spacetime metric field
itself, that should instead only emerge as an approximation of
more fundamental structures in some appropriate limit. In this
sense, the peculiarity of spin foam models as path integrals for
gravity is very attractive: they are constructed out of only
combinatorial and algebraic data, and in a background-independent
fashion, so that no reference to any metric field is needed in
their definition. The algebraic data that are used come from the
local symmetry group of gravity, i.e. the Lorentz group, and the
hope is to be able to describe all the geometry of spacetime and
its dynamics out of this non-geometric information only. However,
it must be said that how this can be possible, in the first place,
and done in practice, is not clear, and more work is certainly
needed.

Now that several spin foam models have been proposed, and that
there is a fairly good understanding of the general formalism, the
question is thus whether or not the proposed models contain the
information needed to reconstruct a metric in some limit and to
recover, in the same limit/approximation, Einstein's equations.
What ingredients does a complete spin foam have to contain? We
know that a classical metric is determined uniquely and almost
completely (one may say to nine tenths) by the knowledge of the
causal structure of spacetime, i.e. the set of causal relations
between points in the manifold, thought of as the conformal
structrure or the set of light cones at each point, with the
remaining degree of freedom being given by a length scale, e.g.
the conformal factor or the determinant of the metric tensor.
Accordingly to this decomposition of the metric degrees of
freedom, one may think of splitting the path integral for quantum
gravity into a sum over causal structures by which we mean both
the set of points and that of their causal relations, and a
sum/integral over possible assignment of scale information, i.e.
metric data which consistently define a length scale, something
like: \be Z\,=\,\sum_C\,\int\,\mathcal{D}l\,\,e^{i\,S} \label{pi}.
\ee

Giving preference to a finitary substitute for continuum
quantities \cite{Sorkin}, the causal structures summed over may be
Lorentzian triangulations, or their topologically dual \lq\lq
Lorentzian 2-complexes", or causal sets, i.e. finitary sets of
points endowed with a partial order representing their causal
relations\cite{Sork}; the length scale, on the other hand, may
also be dealt with in several ways; in spin foam models, which
naturally fit in this finitary approach, both the causal relations
and the length scale have to be encoded in the algebraic data
labelling the 2-complexes. Of course, the amplitude for each
configuration should reflect the causal structure as well.

So we may say that, if we have a spin foam model that defines a
consistent causal structure and length scale, then we can be
pretty sure that it is possible to reconstruct in full a metric
field living in the continuum manifold that one builds up from the
spin foam in some approximation, because all its degrees of
freedom can be uniquely determined.

As for the dynamics of these degrees of freedom, it can be argued
that, if they satisfy, when treated in a statistical mechanics
manner, the thermodynamics governing black holes, in particular
the relation between entropy and area, then the reconstructed
continuum metric will obey Eistein's equation to first
approximation \cite{Jacobson}. It was indeed shown \cite{Jacobson}
that the Einstein equations {\it follow} from these
thermodynamics-geometry relations in a continuum setting. Studying
this statistical mechanics of spin foam degrees of freedom, and
obtaining this relations in a spin foam setting would then be the
next step towards a spin foam quantization of gravity.

In this paper we do not deal with the problem of the dynamics, but
limit ourself to the analysis of the geometric and causal
ingredients of the Barrett-Crane model, leaving the study of the
statistical mechanics of its degrees of freedom to future work.
The discussion above clarifies why we think this is a crucial step
on the path towards a complete quantum gravity theory. Moreover,
such an analysis may help to clarify the relationship between the
spin foam approach and other closely related ones, in particular
dynamical triangulations and causal sets.

\
\
\subsection{Flashback on the path integral realization of the projection
operator and of the Feynman propagator} Given that spin foam
models realise a path integral for quantum gravity, it is still
under question what exactly this algebraic path integral provides,
what kind of transition amplitude, or what kind of Green function
(in analogy with the field theoretic knowledge) it gives.  We
discuss first the path integral or sum-over-histories realization
of Green functions for the relativistic particle, and then how the
same idea can be implemented at least at a formal level in the
case of quantum gravity in the traditional metric (second order)
formalism (most of this section is from \cite{Teitelboim}, but see
also \cite{Halli} for an account of the sum-oev-histories
realization of several Green's functions for the relativistic
particle). Finally, we outline the analogy with the Barrett-Crane
spin foam model, that furnishes the conceptual starting point for
our analysis.

\
\
\subsubsection{Relativistic particle or general relativity on 0 (spatial)
dimensions}
Consider a relativistic particle in flat space, with action:
\be
S(x)\,=\,\int_{\lambda_1}^{\lambda_2}\,(-m)
\sqrt{\frac{dx^\mu}{d\lambda}\frac{dx_\mu}{d\lambda}}\,d\lambda,
\ee
invariant under reparametrization $\lambda\rightarrow f(\lambda)$ (this is
the sense in which one may think
 of this elementary system as similar to general relativity in 0 spatial
dimension) that reduce to the identity
 on the boundary, and its path integral quantization defined by:
\be
Z(x_1,x_2)\,=\,\int_{x_1=x(\lambda_1),x_2=x(\lambda_2)}\,(\prod_{\lambda\in
[\lambda_1,\lambda_2]} d^4 x)\,e^{i\,S(x)}. \ee In order to
understand how different Green's functions come out of this same
expression, as well as to avoid a few technical problems in
dealing with it (see \cite{Teitelboim}), it is convenient to pass
to the Hamiltonian formalism. The action becomes: \be
S(x)\,=\,\int_{\lambda_1}^{\lambda_2}( p_\mu
\dot{x}^\mu\,-\,N\,\mathcal{H})\,d\lambda \ee where
$\mathcal{H}=p_\mu p^\mu + m^2 = 0$ is the Hamiltonian constraint
that gives the dynamics of the system (it represents
 the equation of motion obtained by extremizing the action above with
respect to the variables $x$, $p$, and $N$), and
$p_\mu$ is the momentum conjugate to $x^\mu$.

After a gauge fixing such as, for example, $\dot{N}=0$, one may
proceed to quantization integrating the exponential of the action
with respect to the canonical variables, with a suitable choice of
measure. The integral over the \lq\lq lapse" $N$ requires a bit of
discussion. First of all we use as integration variable
$T=N(\lambda_2 -\lambda_1)$ (which may be interpreted as the
proper time elapsed between the initial and final state). Then
note that the monotonicity of $\lambda$, together with the
continuity of $N$ as a function of $\lambda$ imply that $N$ is
always positive or always negative, so
 that the integration over it may be divided into two disjoint classes
$N>0$ and $N<0$. Now one can show that the integral
over both classes yields the Hadamard Green function:
\bes
G_H(x_1,x_2)\,=\,\langle x_2\mid x_1\rangle &=&
\int_{-\infty}^{+\infty} dT\,
\int (\prod_\lambda d^4 x\, d^4 p)\, e^{i\,\int d\lambda (p x - T
\mathcal{H})} \nonumber \\
&=& \int (\prod_\lambda d^4 x\, d^4 p)\, \delta(p^2\, +\,
m^2)\,e^{i\,\int d\lambda\, (p x)} \ees which is related to the
Wightman functions $G^{\pm}$, in turn obtained from the previous
expression by inserting a $\theta(p^0)$ in the integrand and a
$\pm$ in the exponent, by: \be G_H(x_1,x_2) \, = \,
G^+(x_1,x_2)\,+\,G^-(x_1,x_2)
\,=\,G^+(x_1,x_2)\,+\,G^{+}(x_2,x_1). \label{eq:had} \ee This
function is a solution of the Klein-Gordon equation in both its
arguments, and does not register any order between them, in fact
$G_H (x,y)=G_H (y,x)$. Putting it differently, it is an a-causal
transition amplitude between physical states, or a physical inner
product between them, and the path integral above can be seen as a
definition of the generalized projector operator that project
kinematical states onto solutions of the Hamiltonian constraint,
i.e. as an implementation of the group averaging procedure for
imposing it:

\be
G_H(x_1,x_2)\,=\,\langle x_2\mid x_1\rangle_{phys}\,=\,_{kin}\langle x_2 \mid
\mathcal{P}_{\mathcal{H}=0}\mid x_1\rangle_{kin} .
\ee

On the other hand, one may choose to integrate over only one of the two
classes corresponding to each given sign of $N$, say $N>0$.
This corresponds to an integration over all and only the histories for
which the state $\mid x_2\rangle$ lies in the future
of the state $\mid x_1\rangle$, with respect to the proper time $T$, and
yields the Feynman propagator or causal amplitude:
\be
G_F(x_1,x_2)\,=\,\langle x_2\mid x_1\rangle_C\,=\,\int_{0}^{+\infty} dT\,
\int (\prod_\lambda d^4 x\, d^4 p)\, e^{i\,\int d\lambda (p x - T
\mathcal{H})},
\ee
which is related to the Wightman functions by:
\bes
G_F(x_1,x_2)\,=\,\langle x_2\mid
x_1\rangle_C &=&
\theta(x_1^0\,-\,x_2^0)\,G^+(x_1,x_2)\,+
\,\theta(x_2^0\,-\,x_1^0)\,G^-(x_1,x_2)
\nonumber \\
&=&
\theta(x_1^0\,-\,x_2^0)\,G^+(x_1,x_2)
\,+\,\theta(x_2^0\,-\,x_1^0)\,G^+(x_2,x_1).
\ees

Note that all trajectories of interest, for both orientations of $x_0$
with respect to $\lambda$, are included in the above
integral, if we consider both positive and negative energies, so no
physical limitation is implied by the choice made above.
The resulting function is not a solution of the Klein-Gordon equation, it
is not a realization of a projection onto solutions
 of the Hamiltonian constraint, but it is the physical transition
amplitude between states which takes into account causality
requirements (it corresponds, in field theory, to the time-ordered
2-point function).

\
\
\subsubsection{Quantum gravity in metric formalism (formal)}
Let us now turn to proper quantum gravity, that we deal with in a
rather formal way in the usual metric formalism. The action, in
hamiltonian terms, is: \be S\,=\,\int_\mathcal{M}\,( \pi^{ij}
\dot{g}_{ij}\,-\,N^i\,\mathcal{H}_i\,-\,N\,\mathcal{H} )d^4 x, \ee
where the variables are $g_{ij}$, the 3-metric induced on a
spacelike slice of the manifold $\mathcal{M}$, $\pi^{ij}$, its
conjugate momentum, the shift $N^i$, a Lagrange multiplier that
enforces the (space) diffeomorphism constraint $\mathcal{H}^i=0$,
and the lapse $N$ that enforces the hamiltonian constraint
$\mathcal{H}=0$, again encoding the dynamics of the theory and the
symmetry under time diffeomorphisms, which are required to reduce
to the identity on the boundary, while the space diffeomorphisms
are unrestricted.

We are interested in the transition amplitude between
3-geometries. After a proper gauge fixing (such as $N^i=0$ and
$\dot{N}=0$, \lq\lq proper time gauge"), and the addition of
suitable ghost terms (see \cite{Teitelboim}), one may write a
formal path integral, integrating over the variables with some
given measure the exponential of ($i$ times) the action written
above. The integration over $N^i$ is now absent because of the
gauge condition chosen, but it could be anyway carried out without
problems, with $N^i$ having integration range $(-\infty,
+\infty)$, projecting the states $\mid g_1\rangle$, $\mid
g_2\rangle$ onto diffeomorphism invariant states, while, again, as
in the particle case, the integral over $N$ can be split into two
disjoint classes according to its sign. Again, different choices
of the integration range yield different types of transition
 amplitudes.

The integration over both classes implements the projection onto
physical states, solutions of the Hamiltonian constraint, so that
we may formally write: \bes \langle g_2 \mid g_1\rangle_{phys}
&=&{}_{kin}\langle g_2\mid P_{\mathcal{H}=0}\mid
g_1\rangle_{kin}\,=\,_{kin}\langle g_2\mid
\int_{-\infty}^{+\infty}\mathcal{D}N\,e^{i\,N\,\mathcal{H}}\mid
g_1\rangle_{kin}\nonumber \\
&=&\int_{-\infty}^{+\infty}\,\mathcal{D}N\,\int_{g_1,g_2}
(\prod_{x}\mathcal{D}g_{ij}(x)\mathcal{D}\pi^{ij}(x))\,e^{i\,S},
\ees
where we have omitted the ghost terms.

As in the particle case, this amplitude satisfy all the constraints, i.e.
$\mathcal{H}^\mu \langle g_2 \mid g_1\rangle = 0$,
and does not register any ordering of the two arguments. In this sense it
can be thus identified with the analogue of the
Hadamard function for the gravitational field. It gives the physical inner
product between quantum gravity states.

If we are interested in a physical, causal, transition amplitude between
these states, on the other hand, then we must take
into account the causality requirement that the second 3-geometry lies in
the future of the first, i.e. that the proper time
elapsed between the two is positive. This translates into the restriction
of the integration range of $N$ to positive values
only, or to only half of the possible locations of the final hypersurface
with respect to the first.

Then we define a causal tansition amplitude as:
\bes
\langle g_2 \mid g_1\rangle_C
&=&{}_{kin}\langle g_2\mid \mathcal{E}\mid
g_1\rangle_{kin}\,=\,_{kin}\langle g_2\mid
\int_{0}^{+\infty}\mathcal{D}N\,e^{i\,N\,\mathcal{H}}\mid
g_1\rangle_{kin} \nonumber \\
&=&\int_{0}^{+\infty}\,\mathcal{D}N\,\int_{g_1,g_2}
(\prod_{x}\mathcal{D}g_{ij}(x)\mathcal{D}\pi^{ij}(x))\,e^{i\,S},
\ees where we have formally defined the path integral with the
given boundary states as the action of an evolution operator
$\mathcal{E}$ on kinematical states.

The causal amplitude is not a solution of the hamiltonian constraint, as a
result of the restriction of the average over only
half of the possible deformations of the initial hypersurface, generated
by it; on the other hand, in this way causality is
incorporated directly at the level of the sum-over-histories formulation
of the quantum gravity transition amplitude.

\
\
\subsubsection{Analogy with the BC model}
Coming finally to spin foam models, and to the Barrett-Crane model
in particular, it should be clear that, after having recognised it
as a realization on rigorous grounds of the a path integral for
quantum gravity, one should still answer several key questions:
what kind of amplitude does it define? is it an implementation of
the projector operator or a realization of the Feynman propagator?
if it defines a projector, where is encoded the discrete ${\rm \bf
Z}$ symmetry responsible for the counting of both classes of
\lq\lq lapses" $N$ in the path integral? is it possible to break
such a symmetry, restrict the integration and implement causality
in this way? also, what formulation of spacetime geometry does
 it provide? does it have an intepretation in terms of classical actions
like in \Ref{pi}? what kind of geometric information
is encoded in its amplitudes and how? does it allow the identification of
an underlying finitary causal structure to be
summed over as in \Ref{pi}?

Tentative answers to all these questions will be provided below.

\medskip

A framework for the realisation of the projector resulting from
spin foam models such as the Barrett-Crane model has been proposed
by Perez and Rovelli in \cite{projector}. It is based on the group
field theory underlying the given spin foam model and the
application of the GNS construction to the correlations induced by
the path integral. More precisely, one considers the $C^*$ algebra
of (boundary) spin networks \be {\cal
A}=\left\{a=\sum_sc_ss,\,c_s\in{\mathbf C}\right\} \ee provided
with a natural product defined by the union of two spin networks
$s_1.s_2=s_1\cup s_2$, the star operator defined as $s^*=s$ and
the norm $|a|=\sup_s|c_s|$. Then we consider the state (positive
linear functional) over ${\cal A}$ obtained by a sum of spin foams
$\Delta$ with fixed boundary: \be W(s)=\sum_{\Delta,\,\partial
\Delta=s} A(\Delta), \ee which is defined through the path
integral of the group field theory as \be
W(s)=\int[D\phi]\phi_se^{-S[\phi]}. \ee $\phi_s$ is the function
of the field $\phi$ corresponding to the spin network $s$. For
example, in the Barret-Crane model, the spin networks are
4-valent, the field $\phi$ corresponds to the quantized
tetrahedron and thus creates a 4-valent node: we can glue
4-vertices together to create the spin network $s$, which
corresponds to a product of $\phi$, and $S[\phi]$ is the field
theory action generating the spin foam model through its
perturbative expansion in Feynman diagrams \cite{gft1,gft2}. $W$
is linear by construction and trivially real and positive (the
problematic issue is its convergence). We then simply follow the
GNS construction. Unphysical modes are given by the Gelfand ideal
\be {\cal I}=\left\{a\in{\cal A}|W(a^*a)=0\right\}. \ee Finally
the resulting Hilbert space of physical quantum geometry states is
\be {\cal H}=\overline{{\cal A}/{\cal I}}, \ee with scalar product
given by \be \langle a|b\rangle=W(a^*b). \ee This builds physical
states, solutions of the Hamiltonian constraint. We have started
from the kinematical scalar product
${}_{kin}\langle.|.\rangle_{kin}$, for which spin networks are
orthonormal, to construct the physical scalar product
$\langle.|.\rangle={}_{kin}\langle .|P_{\mathcal{H}=0}|
.\rangle_{kin}= W(.^*.)$. During the process, we have lost time
orientation. This corresponds to the fact that $s^*=s$ and that
consequently the transition amplitude from a state $s_1$ to a
state $s_2$ is real and the same as the reverse phenomenon and as
the process creating $s_1\cup s_2$ out of nothing: \be W(s_1\arr
s_2)=W(s_1s_2)=W(s_1\cup s_2)=W(\emptyset\arr s_1\cup s_2)=
W(s_1\cup s_2\arr\emptyset). \ee The question is then {\it how to
obtain causal transition amplitudes, which are time oriented}. It
seems that requiring $s^*\ne s$ would be enough. This can be
achieved through considering a {\it complex field}. This
possibility is being studied in detail in some work in progress.
In the present paper, we focus on studying the classical and
quantum geometry underlying the Barrett-Crane model and
understanding how it is possible to induce a causal structure in
it, defining causal transition amplitudes. Then, we can conjecture
that a complex group field theory generates this causal spin foam
model just as (real) group field theory generates the present spin
foam models.

\
\
\section{Ponzano-Regge model in dual variables}
Before studying the Barrett-Crane model, linked to general relativity in $3+1$
dimensions, it is interesting to have a look at the corresponding spin foam
model in 3 dimensions. Indeed, in this section, we formulate the Ponzano-Regge model
in a similar fashion as we are going to do with the Barrett-Crane, using normals
to each triangle as basic variables. We explain how the partition function of
the model defines a projector onto the physical states and we show how the model
has a natural $Z_2$ symmetry erasing causality.

\
\
\subsection{Definition of the model}
The Ponzano-Regge model is a simplicial geometry model i.e we view
the (three dimensional) space-time as made from gluing tetrahedra
together and we associate an amplitude to each configuration. More
precisely, we label all edges with an $SU(2)$ spin representation
$V^j$ and give weights to all edges, triangles (faces) and
tetrahedra. The weight associated to edges $e$ is simply the
dimension of the representation $\Delta_j=(2j+1)$. $j$ is the
length of the edge (up to a possible constant). $\Delta_j$ then
corresponds to the number of states $|m\rangle \in V^j$ of an edge
of length $j$. The weight of a triangle $t$ is a function of the
representations $j_1,j_2,j_3$ living on its edges. If there exists
an intertwiner between these three representations (the
Clebsch-Gordan coefficient is non zero), then this weight is 1,
otherwise it is 0. This also corresponds to the number of possible
triangles (to rotations) given the length of its 3 edges.
Mathematically, it is given by the $\Theta$ diagram for the
representations $j_1,j_2,j_3$. It is the product
$\Cm{1}{2}{3}\Cm{1}{2}{3}$ of two Clebsch-Gordan coefficients.
Finally, we associate to each tetrahedron $tet$ the $\{6j\}$
symbol made from the 6 representations living on its edges. By
multiplying all these weights, we get an amplitude associated to a
given configuration of the tetrahedra.

More precisely, the $\{6j\}$ symbol associated to a tetrahedron is
the product of the Clebsch-Gordan coefficients associated to each
of its four triangles:

\be
\{6j\}=\Cm{1}{2}{3}
\Cm{3}{4}{5} \Cm{5}{2}{6} \Cm{6}{4}{1}
\ee

The partition function for a manifold without boundary, equipped with a
fixed triangulation $T$, is then:

\be
Z\,=\,\mathcal{N}_T\left(\prod_e\sum_{j_{e}}\Delta_{j_e}\right)\prod_{t}
\Theta_t(j)\prod_{tet}\{6j\}_{tet} \label{eq:PR}
\ee

where $\mathcal{N}_T$ is a constant depending on the triangulation only.

We can recombine the 3-tensors $C$ associated to each couple of
triangle and tetrahedron to keep them at the level of the
triangle. This is achieved using the following relation (for each
of the two pairs of coefficients living associated to each
triangle, one coming from the theta symbol and the other from one
of the two $6j$-symbols associated to the two tetrahedra sharing
the triangle):

\be
\int_{SU(2)}dh\,D^{j_1}_{a_1b_1}(h)
D^{j_2}_{a_2b_2}(h)D^{j_3}_{a_3b_3}(h)
=\C{1}{2}{3}{a}\C{1}{2}{3}{b}
\ee
where $D^j(h)$ is the representation matrix of the group element $h$
in the representation $j$. This way, we introduce new variables $h$
associated to each triangle of the simplicial manifold. We call these
$SU(2)$ group elements ``{\bf normals} to the triangle''.

Then, the way to construct the Ponzano-Regge amplitude
for a closed manifold
(without boundary), in this ``dual'' formulation, meaning using
these $SU(2)$ variables, is as follows. We use the (combinatorial) 2-complex dual to the
simplicial manifold under consideration, having a vertex for each
tetrahedron, an edge for each triangle and a
plaquette for each edge of the original triangulation. We then divide
each plaquette into ``wedges'', each being the part of the plaquette
``inside'' each 3-simplex (see for example \cite{Reseinberger} for
detailed explanations and pictures).

We put group elements $h$ and $g$ around each wedge inside each
plaquette, i.e. on the links that form the boundary of each wedge.
The $h$ variables are associated to the interior links, there is
one of them for each vertex connected to each interior link, and
can be thought of as associated to the triangles of the simplicial
manifold, as we explained above. The $g$ variables are instead
non-dynamical, in the sense that they do not really contribute to
any amplitude of the model and just disappear if the integration
over them is performed (unless they are located on the boundary of
the 2-complex), and are associated to the links that are shared
between two different wedges (see Fig.1).

\begin{figure}
\begin{center}
\includegraphics[width=8cm]{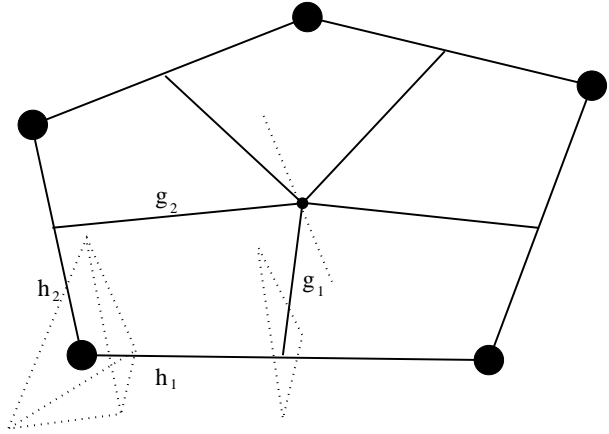}
\caption{A plaquette (dual to an edge of the simplicial manifold)
divided into wedges, with its boundary links (dual to triangles)
and its vertices (dual to tetrahedra) shown, as well as the group
variables assigned to one of its wedges.}
\end{center}
\end{figure}

Then, the amplitude
is the product over the wedges $w$ of the character $\chi^{j_w}$ ($j_w$
labelling a chosen wedge, but the integration over the $g$ variables,
if performed, forces the representations $j$ associated to the
different wedges in the same plaquette to be all the same) evaluated on the oriented product of $g$ and $h$
around the wedge.

The partition function in this ``dual'' formulation (for a manifold
without boundary) then is:

\be
Z\,=\,\mathcal{N}_T\,\left( \prod_w\sum_{j_w}
\Delta_{j_w}\right) \left(\prod_e \int_{SU(2)}dh_e\right)\left(\prod_{\tilde{e}}\int_{SU(2)}dg_{\tilde{e}}\right)\,\prod_v\prod_{w\subset v}\,\chi^{j_w}(h_1g_1g_2h_2)
\ee

where there is an integral for each group element assigned to each
internal link $e$, an integral for each group element assigned to
the boundary $\tilde{e}$ of each wedge within each plaquette, and
a sum over the representations $j$ assigned to each wedge, and the
amplitudes are products of characters for each wedge $w$ inside
each tetrahedron (dual to some vertex $v$). Basically we have been
doing the \lq\lq reverse" of the lattice gauge theory type of
derivation of the Ponzano-Regge model (see \cite{OLoughlin}).
The same kind of ``dual'' formulation for the Euclidean Barrett-Crane model
was shown in \cite{hendryk}.

If we integrate over the $g$ variables, then we just obtain the
constraint that all the representations for the wedges in a given
plaquette are the same, and the measure for it is given by only
one factor $\Delta$ for each plaquette.

If we then sum over the representations $j$, we get the product
over the plaquettes of the Dirac distribution $\delta_{SU(2)}$ evaluated
over the oriented product of $h$ variables only around the boundary of
a full plaquette.

Of course, if we integrate out both the variables $g$ and $h$, we
obtain back the expression \Ref{eq:PR}.

For a manifold with boundary, the boundary is defined by a set of (exterior)
triangles creating a simplicial $2d$ manifold. This results in some
plaquettes being truncated so that some groups elements $g=g_{ext}$
remain exposed on the boundary (they correspond
to a discrete connection living on the boundary). The partition
function is the same as above, but now we do not integrate over the
$g_{ext}$ or do not sum over the $j$ associated with boundary wedges,
depending on which boundary condition we choose to adopt \cite{OLoughlin}\cite{boundary}.
This way, we can construct the amplitude $a(g_{ext},h_{ext})$ of the spin
foam with boundary defined by the boundary connection $g_{ext}$ and the normals
$h_{ext}$ to the triangles on the boundary.

\
\
\subsection{A projection operator}
We construct boundary states, for a given fixed spin foam,
as functionals of both a connection living
on the boundary and  the external ``normals''. They are constructed on
the boundary graph and are functions which therefore depend
on $SU(2)$ group elements $g_e$ living on the edge on the oriented
graph (boundary connection) and on $SU(2)$ group elements $h_v$ living
at the vertices of the graph (triangle normals).
They are required to satisfy the same gauge invariance as the spin
foam amplitude $a(g_{ext}=g_e,h_{ext}=h_v)$:

\be
\forall k_v\in SU(2),\,
\phi(g_e,h_v)=\phi(k_{s(e)}g_ek_{t(e)}^{-1},k_vh_v)
\ee

We can endow this space of functions with
a measure by taking the Haar measure on $SU(2)^E$ where $E$ is the 
number of edges of the graph~: \be \mu(\phi)=\int_{SU(2)^E}
\prod_e{dg_e}\, \phi(g_e,h_v). \ee One can check that this
integral does not depend on the choice of the normals $h_v$ when
$\phi$ is gauge invariant, so that we can choose $\forall
v,\,h_v=Id$ if we want. One can use this measure to define a
Hilbert space of boundary states by considering the space of $L^2$
functions. The resulting scalar product reads

\be
\langle \phi|\psi\rangle=
\int_{SU(2)^E} \prod_e{dg_e}\,
\bar{\phi}(g_e,h_v)\psi(g_e,h_v)
\ee.

Then, one would like to derive the Hilbert space of physical
states taking into account the spin foam amplitude. Following
\cite{projector}, one would like to use the GNS construction to
``project" the space of boundary states down to the physical
states. One considers the algebra ${\cal A}$ of gauge invariant
functions and considers the linear functional~:

\be w(\phi)=\mu(a\phi)=\int_{SU(2)^E}\prod_e{dg_e}\,
a(g_e,h_v)\phi(g_e,h_v) \ee where $a(g_e,h_v)$ is the spin foam
amplitude, summed over the representations, for a boundary
$g_e,h_v$. This functional is positive and we can consider the
corresponding Gelfand ideal~: \be {\cal I}=\{\phi \,|\,
w(\bar{\phi}\phi)=0 \}. \ee Then $w$ defined a norm on ${\cal
A}/{\cal I}$ and the final Hilbert space will be the completion
(for the norm $w$) of ${\cal A}/{\cal I}$.

An easy example is the case of the single tetrahedron (the simplest
triangulation of $B^3$).
The boundary is given by the four triangles (with the topology of $S^2$): we have 4 normals
$h_A,\dots,h_D$ to its four faces and 12 groups elements
$g_1,\dots,g_{12}$ connecting these faces (see Fig.2).

\begin{figure}
\begin{center}
\includegraphics[width=7cm]{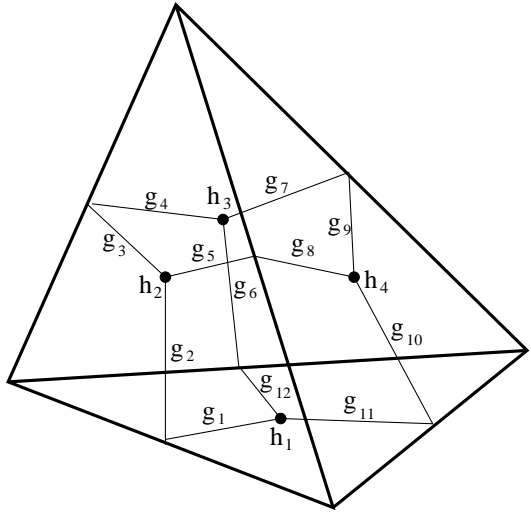}
\caption{A tetrahedron with the boundary dual 1-complex (boundary of
the dual 2-complex) and the relevant connection variables}
\end{center}
\end{figure}

The amplitude can be easily
computed as explained previously and, choosing the gauge $h_v=Id$ to
compute $w$, we get:

\be
w(\phi)=\phi(g_e=Id,h_v=Id).
\ee

Thus only the flat connection $g_e=Id$ is physical, which was the
expected result.

\medskip

That the Ponzano-Regge model is a realization of the projector
operator (which is how we have been using it here) is well-known and was shown, for example, in \cite{Ooguri}
and in \cite{ansdorf}. Let us recall the argument in \cite{Ooguri}.
Consider a 3-manifold $M$ and decompose it into three parts $M_1$,
$M_2$ and $N$, with $N$ having the topology of a cylinder
$\Sigma\times[0,1]$ ($\Sigma$ compact), and the boundaries $\partial M_1$ and $\partial
M_2$ being isomorphic to $\Sigma$. The Ponzano-Regge partition
function may then be written as:

\be
Z_M\,=\,\mathcal{N}_T\,\sum_{j_e\in\Delta_1,j_{\tilde{e}}\in\Delta_2}\,Z_{M_1,\Delta_1}(j_e)\,P_{\Delta_1,\Delta_2}(j_e,j_{\tilde{e}})\,Z_{M_2,\Delta_2}(j_{\tilde{e}})
\ee where we have chosen a triangulation in which no tetrahedron
is shared by any two of the three parts in which we have
partitioned the manifold, and $\Delta_i$ are triangulations of the
boundaries of these parts. We have included also the sum over
spins assigned to edges internal to $M_1$, $M_2$ and $N$ in the
definition of the functions $ Z_{M_i,\Delta_i}$ and
$P_{\Delta_1,\Delta_2}$.

Because of the topological invariance of the model, and of the
consequent invariance under change of triangulation, the following
relation holds:

\be
\mathcal{L}_{\Delta_2}\sum_{j_{\tilde{e}}\in\Delta_2}\,P_{\Delta_1,\Delta_2}(j_e,j_{\tilde{e}})\,P_{\Delta_2,\Delta_3}(j_{\tilde{e}},j'_e)\,=\,P_{\Delta_1,\Delta_3}(j_e,j'_e)
\ee
where $\mathcal{L}$ is another constant depending only on the
triangulation $\Delta_2$.

Because of this property, the following operator acting on spin network
states $\phi_\Delta$ living on the boundary triangulation is a
projector:

\be
\mathcal{P}[\phi_\Delta](j)\,=\,\mathcal{L}_\Delta\,\sum_{j'}\,P_{\Delta,\Delta}(j,j')\,\phi_\Delta(j')
\ee
i.e. it satisfies: $\mathcal{P}\cdot\mathcal{P}=\mathcal{P}$, and we
can re-write the partition function as:

\be
Z_M\,=\,\mathcal{N}_T\,\sum_{j_e\in\Delta_1,j_{\tilde{e}}\in\Delta_2}\,\mathcal{P}[Z_{M_1,\Delta_1}](j_e)\,P_{\Delta_1,\Delta_2}(j_e,j_{\tilde{e}})\,\mathcal{P}[Z_{M_2,\Delta_2}](j_{\tilde{e}}).
\ee

This allows us to define the physical quantum states of the theory
as those satisfying:

\be
\phi_\Delta\,=\,\mathcal{P}[\phi_\Delta] \label{eq:proj}
\ee
being the anlogue of the Wheeler-DeWitt equation,
and the inner product between them as:
\be
\langle \phi_\Delta \mid
\phi'_\Delta\rangle_{phys}\,=\,\sum_{j,j'\in\Delta}\,\phi_\Delta(j)\,P_{\Delta,\Delta}(j,j')\,\phi'_\Delta,
\ee
so that the functions $Z_{M_1}$ and $Z_{M_2}$ are solutions to the
equation \Ref{eq:proj} and the partition function for $M$ basically
gives their inner product.

\medskip
In this argument a crucial role is played by the triangulation
invariance of the model, so that it is not possible to repeat it
for the case of 4-dimensional gravity (i.e. Barrett-Crane model),
where a sum over triangulations or a refining procedure are
necessary to avoid dependence on the given triangulation. However,
it is possible to identify in the Ponzano-Regge model a
distinguishing feature of the projector operator as realized in
path integral terms, and this is the analogue of the $Z_2$
symmetry we have seen in the relativistic particle case and in the
formal path integral quantization of gravity in the metric
formalism. This is the symmetry that ``kills causality'' by
integrating over both signs of the proper time, and is realized in
the present case as a symmetry under change of orientation for the
simplicial manifold (we will give more details on this link
between orientation and causality when discussing the
Barrett-Crane model), as is clear in the Lorentzian context, where
future oriented (d-1)-simplices are changed into past oriented
ones and viceversa. This symmetry is actually evident at the level
of (d-2)-simplices, writing the model in terms of characters as we
did. In fact, under a change of orientation of the edges in the
manifold, the plaquettes (or wedges) of the dual complex also
change their orientation and this is reflected by substituting the
group elements assigned to the boundary links of each plaquette
(wedge) with their inverses. Clearly, the partition function, and
each amplitude in it, is not affected by this change due to the
equality between the characters of group elements which are
inverse of each other. Indeed, in the case of $SU(2)$, a group
element is conjugate to its inverse (the Weyl group is $Z_2$) and
they both have the same (real) character. It is the identification
of an analogous symmetry in the 4-dimensional case that will show
how the Barrett-Crane model realizes a projection onto physical
states. Indeed, once again, $Spin(4)$ and $SL(2,\Cb)$ group
elements are conjugate to their inverse (the Weyl group is still
$Z_2$) and the model is invariant under change of orientation.

In principle, one could try to distinguish the two mirror images
and find a way to restrict the model to only one of them; however,
there is not much motivation to do so in the Euclidean context,
due to the lack of a time direction and all related causality
issues. Then, it would make more sense to look at the same issue
in a Lorentzian $2+1$ spin foam model. Indeed, such a model would
be based on $SU(1,1)$ and its unitary representations. It has been
considered in \cite{davids,2+1}. A very interesting feature of
these constructions is that there exists a topological model based
only on the discrete (positive) series of unitary representations
of $SU(1,1)$. Interestingly, for these representations, it turns
out that the $Z_2$ symmetry is killed (the Weyl group is trivial)
and that the character are simple exponentials: $SU(1,1)$
naturally distinguishes two time arrows, the past and the future.
This model is therefore is a nice candidate for Lorentzian $2+1$
general relativity and represents a motivation to deal with the
$3+1$ case in a similar fashion.

\medskip
The whole previous construction can be generalised to any
dimension. It corresponds to a state sum model for free $BF$
theory. The case which interests us is the 4 dimensional case.
There we are studying a 2-complex locally dual to a 4-dimensional
simplicial manifold i.e a (combinatorial) gluing of 4-simplices.
Each 4-simplex is made of 10 triangles (faces) grouped in 5
tetrahedra. We associate to each triangle a $Spin(4)$
representation in the Euclidean theory (or an $\sl$ representation
in the Lorentzian case). Each tetrahedron is defined as a 4-valent
intertwiner between the 4 representations of its triangles.
$Spin(4)$ can be decomposed as $SU(2)\otimes SU(2)$ and its
irreducible representations are labelled by two half-integers
$J=(j,k)$ for the two corresponding $SU(2)$ representations. A
basis for such a representation is given by the usual basis of
$SU(2)$ representations and is labelled by a couple of
half-integers $M=(m,n)$. There is not a unique 4-valent
intertwiner given four representations, but they form a Hilbert
space. A nice basis is given by decomposing the 4-intertwiner into
two 3-intertwiners. As 3-valent intertwiners are unique (they are
the Clebsch-Gordan coefficients up to normalization), the basis is
labelled by an intermediate $Spin(4)$ representation. We note the
(orthonormalised) intertwiners $\Cq{1}{2}{3}{4}{M}{J}$, where $J$
is the intermediate representation. Finally, the 4-simplex is made
up of 10 triangles labelled by one representation each and 5
tetrahedra also labelled by one representation each. We associate
to each 4-simplex an invariant amplitude made up from these 15
representations. It is the $\{15j\}$ symbol for $Spin(4)$, which
is obtained naturally by contracting the 10 face representations
using the 5 intertwiners. We can give an alternative construction
of the amplitudes. As in the three dimensional case, we associate
$Spin(4)$ (or $\sl$) group elements to each tetrahedron. Then the
relation

\be \int_{Spin(4)} dg\, \prod_{i=1}^4 D^{J_i}_{A_iB_i}(g)=
\Cq{1}{2}{3}{4}{A}{J}\Cq{1}{2}{3}{4}{B}{J} \ee allows us to
reconstruct the whole amplitude of a 2-complex. It gives the
gluing condition of two 4-simplices: the internal representation
of the common tetrahedron is the same for both 4-simplices, as is
obvious from the previous formula. Indeed, the $\Theta$ diagram of
the three-dimensional case is replaced by an {\it eye diagram},
made of four edges labelled with the 4 face representations of the
tetrahedron and with two 4-valent intertwiner (internal)
representations $J$ and $K$ corresponding to the tetrahedron seen
in each 4-simplex. This is the number of quantum states of
tetrahedron. Due to the orthonormality of the $C$ intertwiners,
the eye diagram value is 1 if $J=K$ and 0 else. Also, we have the
same projector interpretation as in the three-dimensional case.

However, we need to stress at this point that the 4-simplex
constructed as such from BF theory is {\bf not geometric}, in the
sense that there is no relationship between the $B$ field of the
theory and any frame or tetrad field, and consequently any metric
field. It is possible to impose such a relationship and also to
relate classical geometric constraints on the structure of a
4-simplex to constraints on the representations \cite{bc1}.  This
gives the Barrett-Crane model which we discuss in the next
section. In that case, the internal representation of a
tetrahedron can be different in the two attached 4-simplices, and
the eye diagram is not restricted anymore to 0 or 1: there is a
richer space of quantum states of the tetrahedron \cite{bb}.

\section{The quantum geometry of the Barrett-Crane model}

Let us now turn to our main object of interest, 4-dimensional
gravity in Minkowskian signature, and to the corresponding spin
foam model: the Lorentzian Barrett-Crane model. We first review
briefly the classical basis for the quantum description of gravity
given by this model, and the model itself; then we show how it
describes the geometry of spacetime both at the quantum and
classical level.

\
\
\subsection{The Lorentzian Barrett-Crane model....}
The classical starting point for the spin foam quantization of
gravity of the Barrett-Crane model is the formulation of
gravitational theory as a constrained topological field theory.
That a final theory of quantum gravity would share the general
properties of a topological field theory was argued many times in
the recent past \cite{Barrett, Crane, mike&carlo}, and that the
implementation of geometric constraints to classical $BF$ theory
reduces this to first order general relativity was know since long
ago \cite{CDJ}; what was missing was a route towards the quantum
translation of these geometric constraints and a framework for
their implementation in a consistent quantum gravity model; this
is provided exactly by the Barrett-Crane model.

Consider the $so(3,1)$-Plebanski action: \be
S(\omega,B,\phi)=\int_{\mathcal{M}}\left[ B^{IJ}\wedge
F_{IJ}(\omega)+\frac{1}{2}\phi_{IJKL}B^{IJ}\wedge B^{KL}\right]
\ee
which is a $BF$ theory (a topological field theory) action with variables
a 2-form $B^{IJ}_{\mu\nu}$
with values in $so(3,1)$, and a 1-form connection $A^{IJ}_{\mu}$ also with
values in the Lorentz
algebra and with curvature $F^{IJ}_{\mu\nu}$,
but with the addition of quadratic constraints on the $B$ field enforced
by the Lagrange multiplier
$\phi_{IJKL}$ (with the symmetry $\phi_{IJKL}\epsilon^{IJKL}=0$).

The equations of motion are: \bes \mathcal{D}B\,=\,dB\,+\,[\omega,B]\,=\,0
\\
 F^{IJ}(\omega)\,=\,\phi^{IJKL}B_{KL} \nonumber \\
B^{IJ}\,\wedge\,B^{KL}\,=\,e\,\epsilon^{IJKL}, \label{eq:Con} \ees and it
can be proven \cite{Mike, DP-F} that the constraints on the $B$ field have
the following four sectors of solutions: \bes
&I&\;\;\;\;\;B^{IJ}\,=\,\pm\,e^{I}\,\wedge\,e^{J} \\
&II&\;\;\;\;B^{IJ}\,=\,\pm\,\frac{1}{2}\,\epsilon^{IJ}\,_{KL}\,e^{K}\,\wedge\,e^{L}
\ees
This means that in one of these sectors: $S\rightarrow S_{EH}=\int
\epsilon_{IJKL}e^{K}\wedge
e^{L}\wedge F^{IJ}$, i.e. the theory reduces to pure Einstein gravity in
first order formalism.

Also, the other sector, differing by a global change of sign only,
is classically equivalent to this, while the other two, related by
Hodge duality to the \lq\lq geometric" ones, corresponds to \lq\lq
pathological geometries" with no physical interpretation (see the
cited literature for more details). It can be shown that,
classically, solutions having their Cauchy data in one sector do
not evolve away from it, unless degeneracy of the tetrad field is
allowed and some pathological case is excluded (see \cite{Mike}
for a complete analysis), so that really the presence of different
sectors of solutions, related to each other by a
$\Z_{2}\times\Z_{2}$ symmetry, may manifest itself physically only
at the quantum level.

In particular, the $\Z_2$ symmetry between the two geometric
sectors of the theory may affect the path integral quantization of
the theory, and the very meaning of the path integral; remember in
fact that a $\Z_2$ symmetry on the lapse function $N\rightarrow
-N$ makes the difference between a path integral realization of
the projector onto solutions of the Hamiltonian constraint and a
path integral representing the Feynman propagator between states,
or causal transition amplitude, both in the relativistic particle
case and in quantum gravity in metric formalism, as we have shown
above. A simple argument suggests that this may happen also in our
spin foam context; in fact, a $3+1$ splitting of the Plebanski
action (see \cite{Peldan}), after the imposition of the
constraints on the $B$ field (we are then analysing the $3+1$
splitting of the Palatini action for gravity), shows that a change
of sign in the $B$ field is equivalent to a change of sign in the
lapse function, so that both sectors of solutions are taken into
account in a path integral realization of the projector operator.
The $B$ field has in fact the role of metric field in this $BF$
type formulation of gravity, and a canonical 3+1 splitting
basically splits its independent components into the triple
($h_{ab}$, $N^a$, $N$) (with all these expressed in terms of the
tetrad field) as in the usual metric formulation. The reason for
and effect of this will be explained below. Of course, more
worrying would be the presence, in the quantum theory, of the two
\lq\lq non geometric" sectors.

A quantization of gravity along such lines should start by identifying
suitable variables corresponding
to the $B$ and $A$ variables of the Plebanski action, and then the correct
translation at the quantum
level of the above constraints on the $B$ field, leading to a realization
of the path integral
\be
\mathcal{Z}\,=\,\sum_\mathcal{M}\int\mathcal{D}B\,\mathcal{D}A\,\mathcal{D}\phi\,\,e^{i\,S(B,A,\phi)}
\ee
(we have included a sum over spacetime manifolds),
possibly in a not only formal way.

Both in light of the \lq\lq finitary" philosophy mentioned above
and hoping to make sense of the path integral by a lattice type of
regularization, we pass to a simplicial setting in which the
continuum manifold is replaced by a simplicial complex, and the
continuum fields by variables assigned to the various elements of
this complex. In particular, the $B$ field is associated to the
triangles in the triangulation by: $B^{IJ}(t)=\int_t
B^{IJ}_{\mu\nu}(x) dx^\mu \wedge dx^\nu$.

With this discretization, the constraints on the $B$ field become
constraints on the bivectors
$B^{IJ}\in so(3,1)\simeq \wedge^2(\Rb^4)$ associated to the various
triangles, and more precisely
\cite{bc1}\cite{bc2}\cite{DP-F}:
\begin{itemize}
\item the bivectors change sign when the orientation of the triangles is
changed;
\item the bivectors are \lq\lq simple", i.e. they satisfy $B(t)\cdot \ast
B(t)=0$;
\item the bivectors associated to neighbouring triangles sum to simple
bivectors,
i.e. $B(t)\cdot \ast B(t')=0$ if $t$ and $t'$ share an edge;
\item the four bivectors associated to the faces of a tetrahedron sum to
zero.
\end{itemize}

The Barrett-Crane quantization then proceeds associating to each
triangle an irreducible representation of the Lorentz group in the
principal unitary series, with the identification:
\be
B^{IJ}(t)\leftrightarrow \ast J^{IJ}((n,\rho)_t),
\ee
where the $J$'s are the generators of the Lorentz algebra, so that a
representation $(n,\rho)_t$ ($n$ a half-integer number, and $\rho$
a real positive number) and the corresponding representation space
(on which the $J$'s act) are attached to each triangle, and
assigning to each tetrahedron a tensor in the space given by the
tensor product of the four representation spaces associated to its
faces. In fact, there exists an ambiguity (Immirzi ambiguity)
in the above correspondence,
which corresponds to identifying $B$ with $\alpha J+\beta *J$.
However, it was shown that such a modification doesn't change the final
resulting spin foam model, which remains the Barrett-Crane model \cite{our}. 
This identification then allows a quantum translation
of the classical constraints given above as:
\begin{itemize}
\item the representations change to their dual when the orientation of the
triangles is changed;
\item the representations to be used are only the \lq\lq simple" ones
$(n,0)$ or $(0,\rho)$, i.e.
those for which $C_2=J((n,\rho)_t)\cdot \ast J((n,\rho)_t)=n\rho=0$, where
$C_2$ is the second
invariant Casimir of the algebra;
\item the representations associated to neighbouring triangles sum to
simple representations,
i.e. $J((n,\rho)_t)\cdot \ast J((n,\rho)_{t'})=0$ if $t$ and $t'$ share an
edge; this also implies that
the tensor for a tetrahedron must be decomposed into its Clebsch-Gordon
series with summands which are
non-zero for simple representations only;
\item the tensor associated to a tetrahedron is an invariant tensor under
the action of $SO(3,1)$.
\end{itemize}

These conditions allow the identification of a unique state $\Psi$ for
each tetrahedron in the triangulation,
given by the invariant tensor described above (the Barrett-Crane
intertwiner); moreover, it can be
shown \cite{bb}, at least in the Euclidean case, that this state
satisfies automatically (i.e.
without further restrictions) the condition:
\be
(U^+ + U^-)\,\Psi\,=\,0
\ee
where $U^\pm$ is the (chiral) operator for the 3-volume of the tetrahedron
\cite{bb}; this fact is
crucial, because it implies that the two non-geometric sectors of
solutions of the classical theory
are in fact {\it absent} in a quantum theory that implements the above
constraints. This is in particular
the case for the Barrett-Crane spin foam model.

To obtain the Barrett-Crane partition function one may proceed in
several ways, using either a lattice gauge theory type of
derivation \cite{danruth, boundary} or the novel technology of
field theories over group manifolds \cite{gft1, ale1, alejandro1,
alejandro2, gft2}. In the first case one starts with a lattice
gauge theory formulation of $BF$ theory and imposes the BC
constraints directly on this using suitable projectors, while in
the second one writes down a partition function for a field living
on four copies of $SL(2,\Cb)$ with a suitable action for it, and
the choice of action leading to the BC model differs from the one
leading to a spin foam formulation of 4-dimensional BF theoy by
the insertion of the same kind of projectors; in the first
approach the simplicial geometry underlying the construction is
slightly more transparent, but one remains limited by a choice of
a fixed triangulation of spacetime, while in the second it is
possible to define a sum over triangulations that gives a complete
formulation of the theory.

In both cases the projectors used realize the simplicity constraint by
imposing that the
representations
of the Lorentz group that label the faces of the 2-complex dual to the
triangulation or, equivalently, the triangles in
the triangulation itself have a realization in the spaces of $L^2$
integrable functions on the hyperboloids in Minkowski
space \cite{bc2}; in particular, the model based on the unitary
irreducible representations $(0,\rho)$ in the principal
series, on which we concentrate our attention here, is obtained imposing $SU(2)$ invariance on the $SL(2,\Cb)$
representation functions corresponding to each face, thus realizing these
representation in the upper hyperboloid
$H^+=\{x\,/\,x\cdot x=1, x^0>0\}\simeq SL(2,\Cb)/SU(2)$ in Minkowski
space; using these representations the third constraint
above regarding the decomposition of the tensor product of simple
representations turns out to be automatically implemented;
the last (closure) constraint is instead implemented as invariance under
the action of the Lorentz group of the product of
the four representation functions corresponding to the four triangles of
the same tetrahedron. In the existing model,
based on real representations of the Lorentz group or, equivalently, on a
real field theory, the first constraint concerning
the orientation of the triangles is not implemented as the final model
does not discriminate between a representation and
its dual; this fact, that can be seen in many different ways, is crucial
for a correct implementation of causality, and for a
correct understanding of the quantum geometry of the model, as we discuss
at length in what follows.

Whatever derivation one decides to pursue, the resulting model,
based on continuous representations, is:
\be
Z\,=\,(\prod_f
\int_{\rho_f}d\rho_f\,\rho_f^2)\,(\prod_{v,
e_v}\int_{H^+}dx_{e_v})\,\prod_e
\mathcal{A}_e(\rho_k)\,\prod_v\mathcal{A}_v (\rho_k, x_i)
\label{eq:Z}
\ee
with the amplitudes for edges (tetrahedra) (to be
considered as part of the measure, as we argue again below) and
vertices (4-simplices) being given by:
\bes
A_e(\rho_1,\rho_2,\rho_3,\rho_4)\,=\,\int_{H^+} dx_1 dx_2
K^{\rho_1}(x_1,x_2)K^{\rho_2}(x_1,x_2)K^{\rho_3}(x_1,x_2)K^{\rho_4}(x_1,x_2)
\\ \\
A_v(\rho_k,
x_i)=K^{\rho_1}(x_1,x_2)K^{\rho_2}(x_2,x_3)K^{\rho_3}(x_3,x_4)K^{\rho_4}(x_4,x_5)K^{\rho_5}(x_1,x_5)
\nonumber \\ K^{\rho_6}(x_1,x_4)
K^{\rho_7}(x_1,x_3)K^{\rho_8}(x_3,x_5)K^{\rho_9}(x_2,x_4)
K^{\rho_{10}}(x_2,x_5) \label{eq:ampl}.
\ees
The variables of the model are thus the representations $\rho$ associated
to the triangles of the
triangulation (faces of the dual 2-complex), there are ten of them for
each 4-simplex as can be seen
from the expression of the vertex amplitude, and the vectors $x_i \in
H^+$, of which there is one for
 each of the five tetrahedra in each 4-simplex amplitude, so that two of
them correspond to each
interior tetrahedron along which two different 4-simplices are glued. We
will discuss the geometric
meaning of all these variables in the following section.
As written, the partition function is divergent, due to the infinite
volume of the Lorentz group (of the hyperboloid), and the immediate
way to avoid this trivial divergence is just to remove one of the
integration over $x$ for each edge and 4-simplex amplitude, or, in
other terms, to gauge fix one variable. It was shown that, after this
gauge fixing is performed, the resulting partition function (for fixed
triangulation) is finite \cite{finite1, finite2}.

The functions $K$ appearing in these amplitudes have the explicit
expression:

\be
K^{\rho_k}(x_i,x_j)=\frac{2\sin(\eta_{ij}\rho_k/2)}{\rho_k\sinh\eta_{ij}}
\label{eq:K} \ee where $\eta_{ij}$  is the hyperbolic distance
between the points $x_i$ and $x_j$ on the hyperboloid $H^+$.

The amplitudes describe an interaction among the $\rho$'s that couples
different 4-simplices and
different tetrahedra in the triangulation whenever they share some
triangle, and an interaction among
the different tetrahedra in each 4-simplex.


In the presence of boundaries, formed by a certain number of
tetrahedra, the partition function above has to be supplemented by
boundary terms given by a function
$C^{\rho_1\rho_2\rho_3\rho_4}_{(j_1k_1)(j_2k_2)(j_3k_3)(j_4k_4)}(x)$
for each boundary tetrahedron \cite{boundary}, being defined as:
\be
C^{\rho_1\rho_2\rho_3\rho_4}_{(j_1k_1)(j_2k_2)(j_3k_3)(j_4k_4)}(x)\,=\,D^{\rho_1}_{00j_1k_1}(x)\,D^{\rho_2}_{00j_2k_2}(x)D^{\rho_3}_{00j_3k_3}(x)\,D^{\rho_4}_{00j_4k_4}(x)
\label{xintertwiner}
\ee
where $x$ is a variable in $\sl/SU(2)$ assigned to each boundary
tetrahedron, $D^{\rho_i}_{00j_ik_i}(x)$ is a representation matrix for
it in the representation $\rho_i$ assigned to the $i-th$ triangle of
the tetrahedron, and the matrix elements refer to the canonical basis
of functions on $SU(2)$, where the $SU(2)$ subgroup of the Lorentz
group chosen is the one that leaves invariant the $x$ vector thought
of as a vector in $\Rb^4$, with basis elements labelled by a
representation $j_i$ of this $SU(2)$ and a label $k_i$ for a vector in
the corresponding representation space.

It is this term that originates the amplitude for internal
tetrahedra after gluing of two 4-simplices along it (it can be
seen as just a half of that amplitude with the integration over
the hyperboloid dropped) \cite{danruth,boundary}.

Also, we stress that the partition function above should be understood
as just a term within some sum over triangulations or over 2-complexes to be defined, for
example, by a group field theory formalism. Only this sum would
restore the full dynamical content of the quantum gravity theory.

\
\
\subsection{..and its quantum and classical geometry.}
Let us now describe the classical and quantum geometry of the
Barrett-Crane model, i.e. the geometric meaning of the
variables appearing in it, the classical picture it furnishes, and the
quantum version of it. Conceptually, the spin foam
formulation of the quantum geometry of a manifold is more fundamental than
the classical approximation of it one uses, and
this classical description should emerge in an appropriate way only in a
semi-classical limit of the model; however, the
derivation of the model from a classical theory helps in understanding the
way it encodes geometric information both at the
classical and quantum level, even before this emergence is properly
understood.

\subsubsection{Geometric meaning of the variables of the model}

In this part, we describe the physical-geometrical content of the
mathematical variables appearing in the model. We summarize and
collect many known facts and show how the resulting
picture that of a discrete piecewise flat
space-time, whose geometry is described by a first order Regge
calculus action (as we then explain in the next section), in order to explain the physical intuition
on which the rest of our work is based.

The variables of the model are the irreducible representations $\rho$'s,
associated to each face of the 2-complex, and the
$x$ variables, associated to each edge, one for each of the two vertices
it links, as we discussed above.

Consider the $\rho$
variables. They result from the assignment of bivector operators to each
face of the 2-complex, in turn coming from the
assignment of bivectors to the triangles dual to them. Given these
bivectors, the classical expression for the area of the
triangles is: $A_t^2=B_t\cdot B_t=B_t^{IJ}B_{tIJ}$, and this, after the
geometric quantization procedure outlined above,
translates into:
$A_t^2=-C_1=-J^{IJ}(\rho_t)J_{IJ}(\rho_t)=\rho_t^2+1>0$\cite{bc1, bc2}.
Thus the $\rho$'s determine the areas of the triangles
of the simplicial manifold dual to the 2-complex. The same result, here
obtained by geometric quantization of the bivector
field, can also be confirmed by a direct canonical analysis of the
resulting quantum theory, studying the spectrum of the
area operator acting on the simple spin network states that constitute the
boundary of the spin foam, and that represent the
quantum states of the theory. We describe all this in the following.
Moreover, from the sign of the (square of the) areas above,
it follows that all the triangles in the model are spacelike, i.e. their
corresponding bivectors are timelike, and consequently
all the tetrahedra of the manifold are spacelike as well, being
constituted by spacelike triangles only\cite{bc2}.
As a confirmation, one may note that we have chosen a particular
intertwiner -the simple or Barrett-Crane intertwiner- between
simple representations such that
it decomposes into only continuous simple representations, which 
translates in algebraic language the fact that timelike bivectors add up
to timelike bivectors.
Clearly, also this sign
property of the areas can be confirmed by a canonical analysis.

Consider now the $x$ variables. They have the natural
interpretation of normals to the tetrahedra of the manifold, and
the tetrahedra being spacelike, they have values in $H^+$, i.e.
they are timelike vectors in $\Rb^4$\cite{bar,bw}. The reason for them being in
$\Rb^4$ is easily explained. The Barrett-Crane model corresponds
to a simplicial manifold, and more precisely to a piecewise flat
manifold, i.e patches (the 4-simplices) of flat space-time glued
together along their common tetrahedra\cite{regge}.
To each flat patch or
4-simplex (a piece of $\Rb^4$) is attached a local reference
frame. In other words, we are using the equivalence principle
replacing the usual space-time points by the 4-simplices: at each
4-simplex, there exists a reference frame in which the space-time
is locally flat. This explains why the normals to the tetrahedra
are vectors in $\Rb^4$ and also why there are two different
normals for each tetrahedron: they are the same vector seen in two
different reference frames. How does the curvature of spacetime
enter the game? Having a curved space-time means that these
reference frames are not identical: we need a non-trivial
connection to rotate from one to another (see also \cite{hendryk}). In this sense, the
non-flatness of spacetime resides in the tetrahedra, in the fact
that there are two normals $x,y\in H^+$ attached to each
tetrahedron, one for each 4-simplex to which the tetrahedron
belongs. The (discrete) connection is uniquely defined,
up to elements of the $SU(2)$ subgroup that leaves invariant
the normal vector on which the connection acts, by the
Lorentz transformation $g$ rotating from one of these normals to
the other ($g\cdot x=y$):
it is a pure boost connection mapping
two points in $H^+$ into one another\footnotemark. Thus, the association of two
variables $x$ and $y$ to each edge (tetrahedron) of the 2-complex
(triangulation) is the association of a connection variable $g$ to
the same edge (tetrahedron) (this is of course the connection variable
associated to the edges, and then constrained by the simplicity
constraints, that is used in all the various derivations of the Barrett-Crane model
\cite{gft1, gft2, ale1, alejandro1, danruth, boundary, hendryk}).
From the product of connection
variables around a closed loop in the dual complex, e.g. the
boundary of a face dual to a triangle, one obtains a measure of
the curvature associated to that triangle, just as in traditional
simplicial formulations of gravity (i.e. Regge calculus) or in
lattice gauge theory. This set of variables, connections on links
and areas on faces, is the discrete analogue of the set of
continuous variables $(B(x), A(x))$ of the Plebanski formulation
of gravity.

\footnotetext{The reconstruction of a (discrete) connection
or parallel transport from the two normals
associated to each tetrahedron was also discussed in \cite{hendryk}
in the context of the Euclidean Barrett-Crane model.
}

Of course, we have a local Lorentz invariance at each ``manifold point" i.e
each
4-simplex. Mathematically, it corresponds to the Lorentz invariance of
the amplitude associated to the 4-simplex. Physically, it says that
the 5 normals $(x_A,x_B,\dots,x_E)$ to the 5 tetrahedra of a given
4-simplex are given up to a global Lorentz transformation:
$(x_A,x_B,\dots,x_E)$ is equivalent to $(g\cdot x_A,g\cdot
x_B,\dots,g\cdot x_E)$. This
is also saying that the local reference frame
associated to each 4-simplex is given up to a Lorentz transformation, as
usual.
Now, given
two adjacent 4-simplices, one can rotate one of the two in order to
get matching normals on the common tetrahedron. If it is possible to
rotate the 4-simplices to get matching normals everywhere, then the
(discrete) connection would be trivial everywhere (i.e. the identity
transformation) and we would get the
equivalent of a classically flat space-time. However, unlike in 3
dimensions where we have only one normal attached to each triangle, in
4 dimensions, we do not want only flat space-times and the ``all
matching normals" configuration is only one particular configuration
among the admissible ones in the Barrett-Crane model.

The presence of such Lorentz invariance shows that the true physical
variables of the model are indeed pairs of $x$ vectors and not
the vectors themselves. One may use this invariance to fix one of the
vectors in each 4-simplex and to express then the
others with respect to this fixed one; in other words the geometric
variables of the model, for each vertex (4-simplex) are
the hyperbolic distances between two vectors corresponding to two
tetrahedra in the 4-simplex, measured in the hyperboloid $H^+$,
i.e. the variables $\eta$ appearing in the formula \Ref{eq:K}. These in
turn have the interpretation, in a simplicial context,
of being the dihedral angles between two tetrahedra sharing a triangle, up
to a sign depending on whether we are dealing with
external or internal angles, in a Lorentzian context (see \cite{BarFoxon}).
These angles may also be seen as the counterpart
of a connection variable inside each 4-simplex.

\medskip
To sum up the classical geometry underlying the Barrett-Crane model,
we have patches (4-simplices) of flat spaces glued together into a curved space.
This curvature is introduced through the change of frame associated
to each patch, which can be identified to the change of time normal
on the common boundary (tetrahedron) of two patches.
Now a last point is the {\it size} of the flat space patches, or in other words
the (space-time) volume of a fixed 4-simplex in term of the 10 $\rho$
representations defining it. This volume can be obtained as the wedge
product of the bivectors associated to two (opposite) triangles -not
sharing a common edge- of the 4-simplex, which reads:
\be
{\cal V}^{(4)}=
\f{1}{30}\sum_{t,t'}\f{1}{4!}
\epsilon_{IJKL}sign(t,t')B_t^{IJ}B_{t'}^{KL}
\ee
where $(t,t')$ are couples of triangles of the 4-simplex and $sign(t,t')$
register their relative orientations.
After quantization, the $B$ field are replaced by the generators $J$ and
this formula becomes:
\be
{\cal V}^{(4)}=
\f{1}{30}\sum_{t,t'}\f{1}{4!}
\epsilon_{IJKL}sign(t,t')J_t^{IJ}J_{t'}^{KL}.
\ee
There is another formula to define the volume, which can be seen as more suitable
in our framework in which we use explicitely the (time) normals to the tetrahedra:
\be
({\cal V}^{(4)})^3=\f{1}{4!}\epsilon^{abcd}
N_a\wedge N_b\wedge N_c\wedge N_d
\ee
where is the oriented normal with norm $|N_i|=v^{(3)}_i$ the
3-volume of the corresponding tetrahedron (more on the 4-volume
operator in the Barrett-Crane model can be found in \cite{danhend}).

\
\
\subsubsection{Simplicial classical theory underlying the model}
\label{area-angle}

Thus the classical counterpart of the quantum geometry of the
Barrett-Crane model, or in other words the classical description
of the geometry of spacetime that one can reconstruct at first from the
data encoded in the spin foam, is a simplicial geometry
described by two sets of variables, both associated to the triangles in the
manifold, being the areas of the triangles
themselves and the dihedral angles between the two normals to the two
tetrahedra sharing each triangle. A classical simplicial
action that makes use of such variables exists and it is given by the
traditional Regge calculus action for gravity. However,
in the traditional second-order formulation of Regge calculus both the
areas and the dihedral angles are thought of as
functions of the edge lengths, which are the truly fundamental variables
of the theory. In the present case, on the other hand, no variable
corresponding to the edge length is present in the model and both the
dihedral angles and the areas of triagles have to be seen as
fundamental variables. Therefore the underlying classical theory for
the Barrett-Crane model is a first order formulation of Regge Calculus
based on angles and areas.

A formulation of 1st order Regge calculus was proposed by Barrett
\cite{balone} in the Euclidean case based on the action: \be
S(l,\theta)\,=\,\sum_t\,A_t(l)\,\epsilon_t\,=\,\sum_t\,A_t(l)\,(
2\pi\,-\,\sum_{\sigma(t)}\,\theta_t(\sigma)) \ee where the areas
$A_t$ of the triangles $t$ are supposed to be functions of the
edge lengths $l$, $\epsilon_t$ is the deficit angle associated to
the triangle $t$ (the simplicial measure of the curvature) and
$\theta_t(\sigma)$ is the dihedral angle associated to the
triangle $t$ in the 4-simplex $\sigma(t)$ containing it. The
dihedral angles $\theta$, being independent variables, are
required to determine, for each 4-simplex, a unique simplicial
metric, a priori different from the one obtained by means of the
edge lengths (actually, unless they satisfy this constraint, the
ten numbers $\theta$'s can not be considered dihedral angles of
any 4-simplex, so we admit a slight language abuse here).
Analytically, this is expressed by the so-called Schl{\"a}fli
identity: \be \sum_t\,A_t\,d\theta_t\,=\,0. \ee The variations of
this action are to be performed constraining the angles to satisfy
such a requirement, and result in a proportionality between the
areas of the triangles computed from the edge lengths and those
computed from the dihedral angles: \be
A_t(l)\,\propto\,A_t(\theta). \ee

The meaning of this constraint is then to assure the agreement of
the geometry determined by the edge lengths and of that determined
by the dihedral angles, and can be considered as the discrete
analogue of the ``compatibility condition'' between the $B$ field
and the connection, basically the metricity condition for the
connection, in the continuum Plebanski formulation of gravity.
Note, however, that this agreement is required to exist at the
level of the areas only. This constraint may also be implemented
using a Lagrange multiplier and thus leaving the variation of the
action unconstrained (see \cite{balone}). In this case, the full
action assumes the form: \be S\,=\,\sum_t\,A_t(l)\,\epsilon_t\,+\,
\sum_\sigma\,\lambda_\sigma\,det\Gamma_{ij}(\theta), \ee where
$\lambda_\sigma$ is a Lagrange multiplier enforcing the mentioned
constraint for each 4-simplex, and the constraint itself is
expressed as the vanishing of the determinant of the matrix
$\Gamma_{ij}=-\cos{\theta_{ij}}=-\cos{(x_i\cdot x_j)}$, where the
$x$'s variables are the normals to the tetrahedra in the 4-simplex
introduced above.

The main difference with the situation in the Barrett-Crane model
is that, in this last case, the areas are not to be considered as
functions of the edge lengths, but independent variables. The
relationship with the dihedral angles, however, remains the same,
and this same first order action is the one to be considered as
somehow ``hidden'' in the spin foam model. We will discuss more
the issue of the simplicial geometry and of the classical action
hidden in the Barrett-Crane model amplitudes, and of its
variation, when dealing with the Lorentzian case.

We
point out that other formulations of first order Regge Calculus exist,
with different (but related) choices for the fundamental variables of
the theory \cite{CDM}\cite{Katsy}\cite{Katsy2}. Also, the idea of using the areas
as fundamental variables was put forward at first in \cite{Rov} and
then studied in
\cite{Make1}\cite{Make2}, with the possibility of describing simplicial
geometry only in terms of areas of triangles, i.e. of inverting the
relation between edge lenghts and areas and thus expressing all
geometric quantities (including the dihedral angles) in terms of the latter, was analysed in \cite{BRW}\cite{MW}\cite{ReggeWill}.

\
\
\subsubsection{Quantum geometry: quantum states on the boundaries and
quantum amplitudes}
From this classical geometry of the Barrett-Crane model, it is easier
to understand the quantum geometry defined in terms of (simple) spin
networks geometry states. Similarly to the Ponzano-Regge case, the
boundary states are Lorentz invariant functionals of both a boundary
connection and the normals on the boundary \cite{boundary2} and a
basis of the resulting Hilbert space is given by the simple spin networks.

More precisely, let's consider a (decorated) spin foam with boundary.
The boundary will be made of 4-valent vertices glued with each other
into an oriented graph. Such structure is dual to a 3d triangulated
manifold. Each edge of the graph is labelled by a (face)
representation $\rho$ and each vertex corresponds to a (simple)
intertwiner between the four incident representations. On each edge $e$,
we can put a group element $g_e$ which will correspond to the boundary
connection \cite{boundary} and we can decompose the simple intertwiner
such that a "normal" $x_v\in{\cal H}_+$ lives at each vertex $v$
\cite{boundary2}.

\begin{figure}
\begin{center}
\includegraphics[width=7cm]{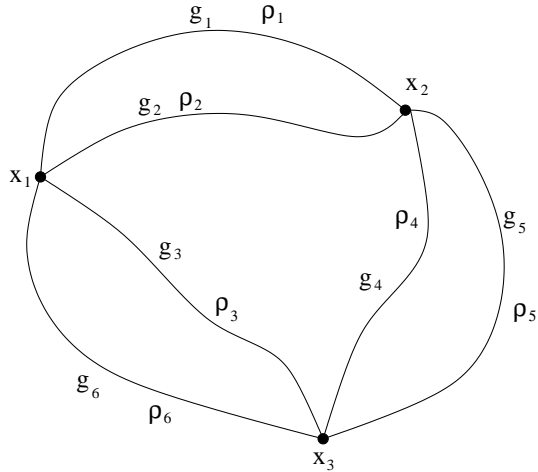}
\caption{A (closed) Lorentz spin network}
\end{center}
\end{figure}

That way, the boundary state is defined by a Lorentz invariant functional
\be
\phi(g_e,x_v)=\phi(k_{s(e)}^{-1}g_ek_{t(e)},k_v.x_v)
\textrm{ for all } k_v\in \sl
\label{gaugeinv}
\ee
where $s(e)$ and $t(e)$ are respectively the source vertex and the
target vertex of the oriented edge $e$. This imposes an $SU(2)$
invariance at each vertex $v$ once we have fixed the normal
$x_v$. One should go further and impose an $SU(2)$ invariance for each
edge incident to the vertex in order to impose the simplicity
constraints \cite{boundary2}.
One endows this space of functionals with the
$\sl$ Haar measure:
\be
\mu(\phi)=\int_{SL(2,\Cb)^E}\textrm{d}g_e\,
\phi(g_e,x_v).
\ee
This measure is independent of the choice for the $x_v$ due to the gauge
invariance \Ref{gaugeinv}.
Then an orthonormal basis of the Hilbert space of $L^2$
functions is given by the simple spin networks~:
\be
s^{\{\rho_e\}}(g_e,x_v)=
\prod_{e}K_{\rho_e}(x_{s(e)},g_e.x_{t(e)})=
\prod_{e}\langle\rho_ex_{s(e)}(j=0)|g_e|\rho_ex_{t(e)}(j=0)\rangle,
\label{simplespinnet}
\ee
where $\mid \rho x(j=0)\rangle$ is the vector of the $\rho$ representation
invariant under $SU(2)_x$ (the $SU(2)$ subgroup leaving the vector $x$
invariant). The general notation is $\mid \rho xjm\rangle$ for the vector
$|m\rangle$ in the $SU(2)$ representation space $V^j$ in the
decomposition $R^\rho=\oplus_j V^j_{(x)}$ of the $SL(2,\Cb)$
representation $\rho$ into representations of $SU(2)_x$.

We should point out that this same Hilbert space for kinematical states
comes out of the canonical analysis of the (generalised)
Hilbert-Palatini action in a explicit covariant framework.
Indeed requiring no anomaly of the diffeomorphism invariance of the
theory, the Dirac brackets, taking into account the second
class constraints (simplicity constraints), of the connection
$\SA_i^X$ ($i$ is the space index and $X$ is the internal Lorentz index)
and the triad $P^i_X$ read \cite{sergei}:
\be
\left\{
\begin{array}{ccc}
\{I\SA,I\SA\}_D &=&0 \\
\{P,P\}_D &=& 0 \\
\{ \SA_i^X,P_Y^j\}_D&=&\delta_i^j I_Y^X.
\label{diracbracket}
\end{array}
\right.
\ee
where $I$ projects the Lorentz index $X$ on its boost part, this
latter being defined relatively to the time normal $x$ (which is built
from the triad field).
One can try to {\it loop quantize} this theory: one would like to
consider spin networks of the connection $\SA$ (cylindrical
functions), but in fact, one needs so-called projected spin networks
which depend on both the connection $\SA$ and the time normal field
\cite{boundary2}.
Then it turns out that
quantizing these commutations relations at a finite number of points
(the vertices $v$ of the graph)
lead to the imposition of the simplicity constraints at the vertices and
the same space of simple spin networks \Ref{simplespinnet} as shown in
\cite{link}.
The graph, representing the quantum geometry state, then has edges
labelled with $SL(2,\Cb)$ simple representations $\rho\ge0$, which
corresponds to an area carried by this edge given by:
\be
area\sim\sqrt{\rho^2+1}.
\label{areaspectrum}
\ee
The restriction to a finite number of points is natural from the spin foam
viewpoint since the space-like slice are made of tetrahedra glued together and
that these same tetrahedra are considered as the points of this 3d-slice.
This explicit relation between the spin foam setting and the canonical theory
is likely to provide us with information on the dynamics of gravity
in both theories.

We have described up to now closed boundaries. What happens is we deal
with an {\it open boundary}? Then one needs to introduce open spin
networks. Let's consider a graph with open ends.
At all vertices will still live a $H^+$ element (time normal).
The edges in the interior will be
defined as previously: on each edge live a $\rho$ representation and a
$\sl$ group element. On the exterior edges, we still have a
$\sl$ representation and a group element, but we introduce some
new label, at the (open) end of the edge, which is a vector in the
(edge) representation. Within its $\sl$ representation, this
vector $v_e$ can be defined in an orthonormal basis
by its $SU(2)$ (sub)representation $j$ and its label
$m$ witin this representation. Then the overall functional is defined
as
\be
s^{\{\rho_e,j_{ext},m_{ext}\}}(g_e,x_v)=
\prod_{e\in int}K_{\rho_e}(x_{s(e)},g_e.x_{t(e)})
\prod_{e\in ext}\langle\rho_ex_{int(e)}(j=0)|
g_e|\rho_e v_e=(j_e,m_e)\rangle,
\ee
where $int(e)$ is the (interior) vertex of the exterior edge $e$.
The scalar product is defined by the integration over all $g_e$
variables. And we get an orthonormal basis labelled by
$\{\rho_e,j_{ext},m_{ext}\}$.

\medskip
We are then ready to explain the geometry behind the amplitude of
the Barrett-Crane model. There are two different ingredients~: the
4-simplex amplitude and the eye diagram corresponding to the
tetrahedron amplitude. The 4-simplex amplitude defines the
dynamics of the space-time; it is to be understood as an operator
going from the Hilbert space of quantum states of a (past)
hypersurface to the Hilbert space of quantum states of (future)
hypersurface deduced from the first one by a Pachner move
corresponding to the 4-simplex. The eye diagram is a statistical
weight which can be understood as the number of quantum states of
a tetrahedron defined by the areas of its 4 triangles. Indeed, a
(classical) tetrahedron is defined by 6 numbers (the 6 lengths of
its edges as an example) and is not fully determined by the 4
areas of its faces: we miss the values of the area of the 2
interior parallelograms \cite{bb}. This indeterminacy persists at
the quantum level. In fact, at the quantum level we can specify,
in addition to the 4 representations corresponding to the 4
triangle areas, an additional variable for each tetrahedron as a
basis for the space of intertwiners between the four
representations in a decomposition of the 4-valent vertex.
However, we are forbidden to specify a further variable
corresponding to the other parallelogram area, by a kind of
indeterminacy relation that forces us to leave it completely
undetermined \cite{bb}. More precisely, let's note $B_1,\dots B_4$
the bivectors of the four triangles of the tetrahedron. Then, one
can construct the areas $A_i=|B_i|$ of the four triangles, the
intermediate areas $A_{ij}=|B_i+B_j|$ of the internal
parallelograms and the volume squared of the tetrahedron
$U=B_1\cdot (B_2\times B_3)$. When quantizing, the bivectors
become $\sl$ generators in some representation $\rho_i$ and
$A_i^2$ become the Casimir of these representations. As for
$A_{ij}^2$, it lives in $\rho_i\otimes\rho_j$ and is diagonal on
each (sub)representation $\rho_{ij}$ in this tensor product. Such
an intermediate representation defines an intertwiner between all
four $\rho_1,\dots,\rho_4$. When computing the commutator of two
such internal areas one gets \be [A_{12}^2,A_{13}^2]=4U. \ee This
way, $A_{12}^2$ and $A_{13}^2$ appear as conjugate variables: if
one fixes the value of $A_{12}$ then $A_{13}$ is undetermined.
This can be understood a purely algebraic level: if one defines
the intertwiner between $\rho_1,\dots,\rho_4$ through $\rho_{12}$,
then this intertwiner decomposes on a whole range of possible
$\rho_{13}$, without fixing it. Thus, the tetrahedron is never
fully specified: the quantum tetrahedron is a fuzzy tetrahedron,
with a fluctuating volume $U$ ($U$ depends on both $\rho_{12}$ and
$\rho_{13}$). Then the eye diagram basically measures the number
of possible states for a tetrahedron (each corresponding to a
possible choice of the unspecified parameter) that do not
correspond to different configuations in the partition function.
This explains why we consider it as part of the measure in the
partition function. This statistical interpretation is
well-defined in the Euclidean case for which the dimension of the
space of states is finite. In the Lorentzian case, the parameters
are continuous and the dimensions become infinite, so that it is
better to think in terms of the connection: we replace the data of
the 2 interior areas by the Lorentz transformation between the
frames of the two 4-simplices sharing the tetrahedron, which
defines the connection internal to the tetrahedron and is
equivalent to the choice of two $H^+$ elements (time-normals). The
{\it eye diagram} can be obtained by integrating over the possible
connections ($\sl$ group elements relating the two normals
attached to the tetrahedron) living at the tetrahedron. This
\lq\lq eye diagram" weight allows all possible connections, thus
it randomizes the connection and correspondingly the space-time
curvature. More precisely, it appears as the {\it amplitude associated
to a connection} living at a tetrahedron defined by its four
triangle representations (areas). We are aware that \cite{baez}
proposed another (natural) interpretation for another tetrahedron
weight in the spin foam amplitude which does not fit with our
point of view. Here, we interpret the eye diagram as a localised
quantum fluctuation corresponding to what we could call a quantum
refractive wave. Refractive waves, as introduced in \cite{refrac},
are a new class of gravitational waves. They allow a discontinuity
in the metric while areas (null-directions) are still
well-defined. This area matching condition is exactly what we have
in the Barrett-Crane model: two touching 4-simplices will agree on
the four face areas of their common tetrahedron while disagreeing
on its complete geometry (different time normals, different
volumes). Moreover it is this indeterminacy which is responsible
for the non-topological character of the Barrett-Crane model.
Indeed, in the topological BF model, as explained previously, the
weight associated to the eye diagram is simply 1.

\medskip
One main difference with the Euclidean Barrett-Crane model is that
the degeneracy of the quantum Lorentzian 4-simplices is
un-probable, so that we would not have to consider any
contribution to the amplitude from such 4-simplices. This is due
to the Plancherel measure. A degenerate 4-simplex would have some
of its triangles of zero area. However, $\rho=0$ has Plancherel
measure $\rho^2=0$ and equivalently has a zero probability. One
can be more precise and look at one particular triangle area
probability graph. As we simply would like to get an idea of
what's happening, to simplify the calculations, one can suppose
the space to be isotropic around the triangle. Then the amplitude
for the area $\rho$ is

\be {\cal A}(\rho)=\rho^2\left(
\f{\sin(\rho\theta)}{\rho\sinh(\theta)} \right)^N \qquad
\theta=\f{1}{N}(2\pi-\theta_0) \ee where $N$ is the number of
4-simplices sharing this same triangle and $\theta_0$ is the
deficit angle corresponding to the curvature around the given
triangle, $\theta_0=0$ corresponding an approximatively flat
space-time. First, the amplitude is 0 for $\rho=0$. Then, we see
that we have a discrete series of maxima which hints toward the
possibility of generating dynamically a discrete area spectrum.
Moreover, this spectrum scales with the curvature. Nevertheless,
there is a big difference of the amplitude of each extremum, which
increases with $N$ i.e when we refine the triangulation.
Therefore, for important $N$, the most probable area $\rho$ is of
order 1 (likely between 1 and 6) and the other extrema are
negligible. This is a reason why it is likely that the relevant
regime will be for plenty of small 4-simplices at small $\rho$ and
not in the asymptotical limit $\rho\arr\infty$.

\medskip
\subsubsection{The Barrett-Crane model as a realization of the
projector operator}
Now we turn to the interpretation of the amplitudes defined by the
Barrett-Crane model. What kind of transition amplitudes do they
represent? We have seen that a path integral realization can be given,
in the relativistic particle case, both for the projector operator
over physical states, thus realizing in a covariant manner the
physical inner product among canonical states, and for the Feynman
propagator, or causal transition amplitude; moreover, we have seen
that a similar situation is present at least formally for quantum
gravity. Now we have a model that furnishes a rigorous construction
for a sum-over-histories formulation of quantum gravity. The issue is
then to realize whether we have in our hands a projector operator or
a causal evolution operator.

We will argue that the Barrett-Crane model is a realization of the
projector operator for quantum gravity. This is not a novel argument
for sure; in fact it was put forward for example in \cite{ansdorf} and
in \cite{projector}.

In \cite{ansdorf}, in fact, a discretization of the projector operator
was proposed, leading to an expression very close to that provided by
spin foam models.

In the 4-dimensional case, for the topology $\Sigma\times [0,1]$, with
$\Sigma$ compact, using a regular lattice with spatial sections $V$
and with time spacing
$\epsilon$ and spatial spacing $l$, this would read:

\bes
P\,:=\,\int\mathcal{D}N\,e^{i\,\int_{0}^{1} dt\int d^3
x\,N(x,t)\,\mathcal{H}(x)}\,
\rightarrow\,
P_{\epsilon,l}\,=\,\int\mathcal{D}N\,
e^{i\,\sum_{t=0}^{1/\epsilon}\sum_{x\in
V}\,\epsilon\,l^3\,N(x)\,\mathcal{H}(x)}\,= \nonumber \\ =
\,\prod_{t}\prod_{x}\,\int_{-\infty}^{+\infty}dT\,e^{i\,\epsilon\,l^{3}\,T\,
\mathcal{H}}\,=\,\prod_{t}\prod_{x}\,\int_{0}^{+\infty}dT\,
\left(
e^{i\,\epsilon\,l^3\,T\,\mathcal{H}}\,+\,e^{-i\,\epsilon\,l^3\,T\,\mathcal{H}}
\right)\,=\,\prod_{t}\prod_{x}\,U_{\epsilon,l}(t,x).
\ees

The expression above is then interpreted as a sum over triangulations
and the operators $U_{\epsilon,l}$, being (a sum over) local evolution operators,
implement the action of the projector operator on a given spin
network and are identified with the quantum operators corresponding to
given Pachner moves in the spin foam case \cite{ansdorf}. The integral over $T$ is
then the analogue of the integral over the algebraic variables in the
spin foam model.

More generally, the idea is, we stress it again, to regard the
partition function for a spin foam model as giving the physical
inner product between (simple) spin network states as the action
of a projector operator over kinematical spin network states: \be
\langle \Psi_2 \mid \Psi_1 \rangle_{phys}\,=\,\langle \Psi_2 \mid
\mathcal{P}\mid \Psi_1
\rangle_{kin}\,=\,\sum_\Delta\,\lambda(\Delta)\,Z_\Delta, \ee
where the full partition function for the given spin foam model
has been split into a sum over triangulations (or their dual
2-complexes) $\Delta$ of partition functions associated with each
given one.

Still at a formal level, one may thus give a discretized
expression for the projector operator as a sum over triangulations
of projectors associated with each triangulation. In general one
may expect then this projector to have the form: \bes
Z(\Psi_1,\Psi_2)&=&_{kin}\langle \Psi_2\mid P_{\mathcal{H}=0}\mid
\Psi_1\rangle_{kin}=\int_{-\infty}^{+\infty}dT e^{i\,T\,\int
dx\mathcal{H}(x)}
=\sum_\Delta\lambda(\Delta)\sum_\sigma\int_{-\infty}^{+\infty}dT_\sigma
e^{i\,T_\sigma\,\mathcal{H}_\sigma}= \nonumber\\
&=&\sum_\Delta\lambda(\Delta)\sum_\sigma\sum_{t_\sigma}\int_{-\infty}^{+\infty}dT_{t_\sigma}
e^{i\,T_{t_\sigma}\,\mathcal{H}_{t_\sigma}}\,=\,
\sum_\Delta\lambda(\Delta)\sum_\sigma\sum_{t_\sigma}\int_{0}^{+\infty}dT_{t_\sigma}\left(
e^{i\,T_{t_\sigma}\,\mathcal{H}_{t_\sigma}}\,+\,
e^{-\,i\,T_{t_\sigma}\,\mathcal{H}_{t_\sigma}}\right)\,\,\,\,\,\,\,\,\,
\label{eq:projdiscr},
\ees
depending on whether the discretization procedure associates the
relevant integral (i.e. the relevant integration variable $T$ or some
analogue of it) to the
4-simplices $\sigma$ or more specifically to each triangle $t_\sigma$ within
each 4-simplex $\sigma$.

To obtain a more conclusive argument for considering a given spin foam
model as realizing the projection over physical states one has however
to go beyond this and try to identify the distinguishing features of
the projector operator in the particular spin foam model under
consideration.

What are then these distinguishing features?

We have already discussed how the difference between a path
integral realization of the projector operator and a path integral
formula for the causal transition amplitude is marked uniquely by
the presence in the former of a $Z_2$ symmetry relating positive
and negative lapses (or proper times), both in the relativistic
particle case and in the formal path integral quantization of
gravity in the metric formalism.

We have also seen this symmetry present in the Ponzano-Regge
model, which is the 3-dimensional model corresponding to the
4-dimensional Barrett-Crane one.

Moreover, we have also noted that a similar and related $Z_2$
symmetry has to be present in the Barrett-Crane model as well as a
result of the quantization procedure that originated its
construction.

The task is then to locate clearly where, in the peculiar structure of
the amplitudes of the model, this symmetry is implemented and how.

Let us come back then to the expression for the Barrett-Crane
partition function \Ref{eq:Z}.

We see that, considering the amplitude
for the faces and those for the internal edges as part of the measure,
as we suggest is the most reasonable thing to do, the model has
exactly the form \Ref{eq:projdiscr}, since the amplitude for the vertices
of the 2-complex, encoding the dynamics of the theory and thus related
to the hamiltonian constraint, is given by a product of terms each
associated to the triangles in the 4-simplex dual to the given vertex.

Even more remarkably, the $K$ functions associated with each
triangle are given by a sum of two exponentials with opposite
signs (with weighting factors in the denominators), resembling the
structure of the formal discretization given above.

Can we interpret the $K$ functions, because of their structure, as encoding the
same $Z_2$ symmetry in the proper time (or in the lapse) that
characterizes the projector operator?
The answer is yes.

Consider in fact that, as we noted above, the symmetry
$T\rightarrow -T$ in the proper time is the same as the symmetry
$N\rightarrow -N$ in the lapse function, in turn originating from
the symmetry $B\rightarrow -B$ in the bivector field, that we
expect to be present in the model at the quantum level. A change
in the sign of $B$ in our discretized context corresponds to a
change in the orientation of the triangles on the simplicial
manifold, and to a consequent change in the orientation of the
tetrahedra. Therefore we expect the sum over both signs of proper
time, assuring the implementation of the Hamiltonian constraint
but also erasing causality requirements from the model, to be
realized in our simplicial model as a sum over both orientations
of the triangles and of the tetrahedra in the manifold. This is
indeed what happens.

In fact, let us analyse more closely the $Z_2$ symmetry of the $K$
functions.
We use the {\it unique} decomposition of $\sl$
representation functions of the 1st kind (our $K$ functions) into representation
functions of the 2nd kind (we denote them $K^{\pm}$)\cite{Ruhl} and we write the $K$ function as
\bes
K^{\rho_{ij}}(x_i,x_j)&=&\frac{2\sin(\eta_{ij}\rho_{ij}/2)}{\rho_{ij}\sinh\eta_{ij}}
=\frac{e^{i\,\eta_{ij}\,\rho_{ij}/2}}{i\rho_{ij}\sinh\eta_{ij}}\,-
\,\frac{e^{-\,i\,\eta_{ij}\,\rho_{ij}/2}}{i\rho_{ij}\sinh\eta_{ij}}\,
\nonumber \\
&=&K^{\rho_{ij}}_{+}(x_i,x_j)+K^{\rho_{ij}}_{-}(x_i,x_j)=
K^{\rho_{ij}}_{+}(x_i,x_j)+K^{\rho_{ij}}_{+}(-x_i,-x_j)
\nonumber \\
&=&K^{\rho_{ij}}_{+}(x_i,x_j)+K^{\rho_{ij}}_{+}(x_j,x_i)=
K^{\rho_{ij}}_{+}(\eta_{ij})+K^{\rho_{ij}}_{+}(-\eta_{ij}),
\label{eq:Ksplit}
\ees
which makes the following alternative expressions
of the same $Z_2$ symmetry manifest: \be
K^{\rho_{ij}}(x_i,x_j)\,=\,K^{\rho_{ij}}(\eta_{ij})\,=
\,K^{\rho_{ij}}(-\eta_{ij})\,=\,K^{\rho_{ij}}(-x_i,-x_j)\,=
\,K^{\rho_{ij}}(x_j,x_i). \ee We see that the symmetry
characterizing the projector operator is indeed implemented as a
symmetry under the exchange of the arguments of the $K$ functions
associated to each triangle in each 4-simplex (so that the
resulting model does not register any ordering among the
tetrahedra in each 4-simplex), or as a symmetry under the change
of sign (orientation) of the two tetrahedra sharing the given
triangle (so that the resulting model does not register the
orientation of the triangle itself), and these in turn are
equivalent to a change in sign of the hyperbolic distance between
the two points on the upper hyperboloid $H^+$ identified by the
vectors normal to the tetrahedra themselves. It is also possible
to consider negative hyperbolic distance on the upper hyperboloid
$H^+$ as positive distances on the lower hyperboloid $H^-=\{
x\in\Rb^{3,1} | x\cdot x = 1, x^0 < 0\}$ in Minkowski space, so
that the symmetry we are considering is in the way the model uses
both the upper and lower hyperboloids in Minkowski space. Note
also the analogy between \Ref{eq:Ksplit}, for the kernels $K$ and
\Ref{eq:had} for the Hadamard Green function for the relativistic
particle (maybe it is possible to make this analogy more precise
by studying the quantization of relativistic particles on the
upper hyperboloid $H^+$, but we do not analyze this possibility in
the present paper).

It is interesting to note also that the reality and positivity of
the partition fuction \cite{BaezChristensen}, a puzzling feature
indeed if the model was to represent a sum-over-geometries
realization of a causal transition amplitude, i.e. the matrix
elements in a spin network basis of some evolution operator, is
perfectly understandable if the model represents a path integral
realization of the projector operator (it is always possible to
find a basis such that the matrix elements of the projector
operator are real and positive). This symmetry can be traced back
exactly to the reality of the $K$ functions and this in turn can
be clearly seen as due to the symmetry worked out above.

\section{Implementing causality: a causal transition amplitude  and the causal set reformulation
of the BC model}

We turn now to the problem of implementing causality in the
Barrett-Crane model, i.e. constructing a causal transition
amplitude starting from the a-causal one. Having identified the
$Z_2$ symmetry that erase causality from the model, the problem is
now to break this symmetry in a consistent and meaningful way. Our
strategy is to find this consistent restriction by requiring that
the resulting amplitude has stationary points corresponding to
good simplicial Lorentzian geometries. We analyse first the case
of a single 4-simplex and give at the end a causal amplitude for
the whole simplicial manifold. After we have obtained the causal
amplitude, we show how the resulting causal model fits in the
general framework of quantum causal sets (or histories).

\subsection{A causal amplitude for the Barrett-Crane model}

\subsubsection{Lorentzian simplicial geometry}
First, we will carry out a stationary point analysis in the
Lorentzian setting as done in the Euclidean case in \cite{bw},
more precisely we study the action associated to a single
4-simplex and study its (non-degenerate) stationary point. They
will turn out to be oriented Lorentzian 4-simplices\footnotemark
\footnotetext{During the course of writing the present manuscript,
the results reported in this subsection and in the following were
published in \cite{schlafli}}.

For this purpose, we first give some details about simplicial
Lorentzian geometry, introduce notations and derive the
Schl{\"a}fli identity encoding the constraint between the angle
variables in a first order Regge formalism based on the area-angle
variables as explained in section \ref{area-angle}. We denote by
$x_i\in{\cal H}^+$ the five un-oriented normals to the tetrahedra
of the 4-simplex and $n_i=\alpha_ix_i$ the oriented (outward)
normal with $\alpha_i=\pm 1$ (we can think also of $+x_i$ as
living in $\mathcal{H}^+$ and of $-x_i$ as living in $\cal{H}^-$).
We call {\it boost parameter}, associated to the triangle shared
by the $i$ and $j$ tetrahedra, the quantity
\be
\eta_{ij}=\cosh^{-1}(x_i\cdot x_j)\ge 0 \label{angledef}
\ee
and then the {\it dihedral Lorentz angle} is defined as
\be
\theta_{ij}=\alpha_i\alpha_j\eta_{ij}=
\alpha_i\alpha_j\cosh^{-1}(\alpha_i\alpha_jn_i.n_j).
\ee
When $\alpha_i\alpha_j=1$, $\theta_{ij}=\eta_{ij}$ is called an
interior angle and, in the opposite case, it is called an exterior
angle. Then, the (spherical) kernel in the $\rho$ representation
reads
\be
K_\rho(x_i,x_j)=\f{\sin\rho\eta_{ij}}{\rho\sinh\eta_{ij}}
=\sum_{\epsilon=\pm1} \f{\epsilon}{i\rho\sinh\eta_{ij}}
e^{i\epsilon\rho\eta_{ij}}. \ee

The Schl{\"a}fli identity is the (differential of the)
constraint relating the 10 angles stating that they come from the
5 normals to the tetrahedra: the 5 normals $n_i$ defined a unique
(geometric) 4-simplex (up to overall scale) \cite{bc1} and the 10
angles defined from them through \Ref{angledef} are therefore not
independent. They satisfy
\be
\sum_{i\ne j} 
A_{ij}\textrm{d}\theta_{ij}
=\sum_{i\ne j} \alpha_i\alpha_j
A_{ij}\textrm{d}\eta_{ij}
=0
\label{schlaflieq} \ee
where $A_{ij}$ are the areas of the 10 triangles of the defined
4-simplex. Following \cite{balone}, the proof is straightforward
from the definition of the angles. Let us call
$$
\gamma_{ij}=n_in_j=\alpha_i\alpha_j\cosh\eta_{ij}
=\alpha_i\alpha_j\cosh(\alpha_i\alpha_j\theta_{ij}).
$$
Then the closure of the 4-simplex reads \be
\sum_{i=1}^{i=5}|v^{(3)}_i|n_i=0 \ee where $v^{(3)}_i$ is the
3-volume of the tetrahedron $i$. This implies \be \sum_i
|v^{(3)}_i|\gamma_{ij}=0,\;\;\;\forall j, \ee that is the
existence of the null vector $(|v^{(3)}_i|)_{i=1\dots5}$ for the
matrix $\gamma_{ij}$. Differentiating this relation with respect
to the metric information (this will be the area variables in our
case) and contracting with the null vector, we obtain \be
\sum_{i\ne j} |v^{(3)}_i||v^{(3)}_j|\textrm{d}\gamma_{ij}=
\sum_{i\ne j} |v^{(3)}_i||v^{(3)}_j|\alpha_i\alpha_j
\sinh(\eta_{ij})\textrm{d}\eta_{ij} =0. \ee
Finally, one can easily show that \be
\sinh(\eta_{ij})|v^{(3)}_i||v^{(3)}_j|=
\f{4}{3}|{\cal V}^{(4)}|A_{ij} \ee where ${\cal V}^{(4)}$ is the
4-volume of the 4-simplex. Then one can conclude with the
Schl{\"a}fli identity \Ref{schlaflieq}.

The first order action for Lorentzian Regge calculus, with the
constraint on the dihedral angles enforced by means of a Lagrange
multiplier $\mu$ \cite{balone}, then reads:

\be
S_{R}\,=\,\sum_t\,A_t\,\epsilon_t\,+\,\sum_\sigma\,\mu_\sigma\,det(\gamma^\sigma_{ij})\,=\,\sum_t\,A_t\,\sum_{\sigma/t\in\sigma}\,\theta_{ij}(\sigma)\,+\,\sum_\sigma\,\mu_\sigma\,det(\gamma^\sigma_{ij})
\ee
where the areas and the angles are considered as independent variables.

\subsubsection{Stationary point analysis and consistency conditions on
the orientation}
Let us now turn to the Barrett-Crane amplitude and study its stationary points.
For a given {\bf fixed} triangulation $\Delta$, with triangles $t$, tetrahedra $T$
and 4-simplices $s$, the amplitude reads
\be
A(\Delta)=\sum_{\epsilon_t=\pm1}\int\prod_t\rho_t^2\textrm{d}\rho_t
\prod_T A_{eye}(\{\rho_t,t\in T\})
\prod_s\int_{({\cal H}^+)^4} \prod_{T\in s}\textrm{d}x^{(s)}_T
\left(\prod_{t\in s}
\f{\epsilon_t}{i\rho_t\sinh\eta_t}\right)
e^{i\sum_{t\in s}\epsilon_t\rho_t\eta_t}
\ee
where $\rho_t$ are the representation labelling the triangles (faces of the
triangulations, dual to the faces of the 2-complex) and $A_{eye}(T)$ are the
amplitude of the eye diagram associated to the tetrahedron $T$
(obtained by gluing together two 4-intertwiners labelled by the 4 representations
living on the faces of $T$).

The action for a single (decoupled) 4-simplex is then \be
S=\sum_{t\in s}\epsilon_t\rho_t\eta_t, \ee with the angles
$\theta_t$ given by \Ref{angledef} and thus constrained by the
Schl{\"a}fli identity. Consequently, we have \be
\textrm{d}S=\sum_{t=(ij)\in s}\epsilon_t\rho_t\textrm{d}\eta_t
=\mu\times \sum_{i\ne j} \alpha_i\alpha_j
A_{ij}\textrm{d}\eta_{ij} \ee where $\mu\in\mathbf{R}$ is the
Lagrange multiplyer enforcing the constraint. Therefore, we obtain
the following equations defining the stationary points \be \left\{
\begin{array}{ccc}
\epsilon_{ij}\alpha_i\alpha_j&=&sign(\mu) \\
\rho_{ij}&=&|\mu|A_{ij}
\end{array}
\right.
\label{localorient}
\ee
This means that the area of the triangles are given (up to scale) by the
$\sl$ representation labels $\rho_{ij}$ and that we have a consistency
relation between the orientation of the tetrahedra $\alpha_i$, the orientation
of the triangles $\epsilon_{ij}$ and the global orientation
of the 4-simplex $sign(\mu)$.
This means that only some particular choices of values for $\epsilon$
corresponds to stationary points being represented by well-defined
Lorentzian simplicial geometries

These stationary oriented 4-simplices are supposed to be the main contribution
to the full path integral in the asymptotical limit $\rho_{ij}\rightarrow\infty$
up to degenerate 4-simplices as dealt with in
\cite{baez2,schlafli,laurentasymp}. Indeed, such degenerate 4-simplices
dominate the (standard) Euclidean Barrett-Crane model in the asymptotical limit.
Ways to sidestep this problems have been proposed through
modifications of the model
in \cite{baez2,laurentasymp}. However, it will maybe turn
out that the asymptotical
limit is not the physically relevant sector and that small
$\rho$'s would dominate the
path integral, when we take into account the coupling between
4-simplices (through the representations $\rho$'s).
Nevertheless, we will not discuss this issue in the present work and
we will focus on the orientation issue and the resulting causal structure of the
discrete spin foam manifold.

\medskip

Through the stationary point analysis, we have analysed the
orientability of a single quantum 4-simplex, and obtained consistency
relations between the global oriention of the simplex, the
orientations of its 5 tetrahedra and the orientations of its 10
triangles (linking the tetrahedra).
Now, can we extend this orientation to the whole spin foam~?
This means having consistent orientations of all the
tetrahedra: if a tetrahedron is past-oriented for one 4-simplex,
then it ought to be future-oriented for the other.
Therefore a consistent orientation is a choice of $\mu_v$ (for
4-simplices) and $\alpha_{T,v}$ (for each tetrahedron $T$ attached to a
4-simplex $v$) such that:

\be
\forall T,\,\mu_{p(T)}\alpha_{T,p(T)}=-\mu_{f(T)}\alpha_{T,f(T)}
\label{globalorient}
\ee
where $v=p(T)$ and $v=f(T)$ are the two 4-simplices sharing $T$.

In fact, this imposes constraints on the signs around each loop of
4-simplices which are equivalent to requiring an orientable
2-complex and \cite{gft1} proposed a way to generate only
orientable 2-complexes from the group field theory defining the
spin foam model. Indeed the field over $\sl^4$ (or $Spin(4)$ in
the Euclidean case), which represents the quantum tetrahedron, is
usually chosen to be symmetric under any exchange of its
arguments:
\be
\phi(g_1,g_2,g_3,g_4)=\phi(g_{\sigma(1)},g_{\sigma(2)},
g_{\sigma(3)},g_{\sigma(4)})
\ee
It was proposed to orient the tetrahedron by requiring its invariance
under even permutations $\epsilon(\sigma)=1$. This was shown to lead
to only orientable spin foams.

\subsubsection{A causal transition amplitude}
Now that we know what are the consistency conditions on the
$\epsilon$ that correspond to well-defined geometric
configurations, we can fix these variables and thus break the
$Z_2$ symmetry they encode (average over all possible orientations
of the triangles) and that erases causality from the model. This
means choosing a consistent orientation
$\{\mu_v,\alpha_{T,v},\epsilon_t\}$ for all simplices satisfying
\Ref{localorient} and \Ref{globalorient}. In other words we fix
the $\epsilon_t$ in the amplitude, and thus the orientation, to
the values corresponding to the stationary points in the a-causal
amplitude (of course we do not impose the other conditions coming
from the stationary point analysis, i.e. we do ot impose the
equations of motion). Then, this leads to a {\it causal
amplitude}, constructed by picking from the general Barrett-Crane
model only the terms corresponding to the chosen orientations
$\epsilon_t$ of the triangles:

\bes
A_{causal}(\Delta)&=&\prod_s\int_{({\cal H}_+)^4}
\prod_{T\in s}\textrm{d}x^{(s)}_T
\prod_{t\in s}
\f{\epsilon_t}{i\rho_t\sinh\eta_t}\int\prod_t\rho_t^2\textrm{d}\rho_t
\prod_T A^T_{eye}(\{\rho_t\}_{t\in T})
\prod_s\,e^{i\sum_{t\in s}\epsilon_t\rho_t\eta_t} \nonumber\\
&=&\prod_s\int_{({\cal H}_+)^4} \prod_{T\in s}\textrm{d}x^{(s)}_T
\prod_{t\in s}
\f{\epsilon_t}{i\rho_t\sinh\eta_t}\int\prod_t\rho_t^2\textrm{d}\rho_t
\prod_T A^T_{eye}(\{\rho_t\}_{t\in T})\,e^{i\,\sum_t\,\rho_t\,\sum_{s|t\in
s}\theta_t(s)}\nonumber\\
&=&\prod_s\int_{({\cal H}_+)^4} \prod_{T\in s}\textrm{d}x^{(s)}_T
\prod_{t\in s}
\f{\epsilon_t}{i\rho_t\sinh\eta_t}\int\prod_t\rho_t^2\textrm{d}\rho_t
\prod_T A^T_{eye}(\{\rho_t\}_{t\in T})\,e^{i\,S_{R}}
\label{orientampli}
\ees
For a (simplicial) manifold $\Delta$ with boundaries, we need to take in
account the boundary term, which are exactly the same as for the usual
Barrett-Crane model.
This corresponds to a particular local causal structure defined by the
relative orientation of the 4-simplices: a 4-simplex is in the
immediate past of another if they share a tetrahedron whose
normal is future oriented for the first and past oriented for the
other.

Of course the above amplitude has to be understood within a sum over
oriented 2-complexes or triangulations:

\be
Z_{causal}\,=\,\sum_\Delta\,\lambda(\Delta)\,A_{causal}(\Delta).
\ee

An interesting question is: do we have
any closed time-like curves? Taking all $\mu$ positive, these
time-like curves correspond to a loop (closed sequence of 4-simplices
linked by common tetrahedra) along which all triangles have a
$\epsilon_t=-1$ orientation. A priori such defects are allowed.
Maybe they would automatically vanish when looking at the stationary points of the
full spin foam amplitude.
Nevertheless, we can take advantage of the fact that in the present
context we are working with a fixed triangulation, and just consider only
oriented triangulation without such patological configurations
allowed.
This way, we get a proper (global) causal
structure on the spin foam. The Barrett-Crane model can then be
considered not simply as a sum over (combinatorial) manifolds $\Delta$
but more precisely as a sum over manifolds
provided with a (consistent) causal structure
$(\Delta,\{\mu_v,\alpha_T,\epsilon_t\})$.

\medskip
Let us now discuss the features of the causal amplitude, and compare
the resulting spin foam model with other approaches to Lorentzian quantum gravity.

Let us first re-write the causal partition function in a
simplified (and physically more transparent) form. Basically, it is just:

\be
Z_{causal}\,=\,\sum_\Delta\,\lambda(\Delta)\,
\int\mathcal{D}\theta_t(\Delta)\,
\int\mathcal{D}A_t(\Delta)\,e^{i\,S_{R}^\Delta(A_t,\theta_t)}
\ee

What are its features?
\begin{itemize}
\item It is clearly a realization of the sum-over-geometries approach
to quantum gravity, in a simplicial setting; its ingredients are: a
sum over causally well-behaved triangulations (or equivalently
2-complexes), where the causal relations are encoded in the
orientation of each triangulation (2-complex); a sum over metric
data for each triangulation, being given by the areas of triangles and
dihedral angles, treated as independent variables, and with a {\it
precise assignment of a measure} (albeit a rather complicated one) for
them; an amplitude for each geometric configuration given by the
exponential of ($i$ times) the first order Lorentzian Regge action. It
can thus be seen as a path integral for Regge calculus, with an
additional sum over triangulations.

\item The geometric variables have a natural algebraic characterization and
origin in the representation theory of the Lorentz group, and the only
combinatorial data used from the underlying triangulation is from the
``first 2 levels'' of it, i.e. from its dual 2-complex only; in other
words, the model is a spin foam model\cite{baezspinfoam}\cite{danrev}.

\item It realises the general definition given in \cite{gupta} for a causal spin foam model (it is, to the best of our
knowledge, its first non-trivial example), except for the use of
the full Lorentz group instead of its $SU(2)$ subgroup; its
boundary states are thus covariant (simple) spin networks and it
can be regarded as giving a path integral definition of the
Hamiltonian constraint in covariant loop quantum gravity.

\item Each ``first layer'' of the dual 2-complex being a causal set,
the model identifies it a the fundamental discrete structure on which
quantum gravity has to be based, as in the causal set approach
\cite{Sork}; the main difference is that it contains additional metric
data intended to determine a consistent length scale (volume element),
which instead, in the traditional causal set approach is meant to be
obtained by ``counting only'', i.e. just in terms of the combinatorial
structure of the causal set (number of vertices, etc.); this is a
particular case included in the general form of the partition function
above; in fact, if we fix all the geometric data to some arbitrary
value (and thus neglect the integrals in the formula), then presumably
any calculation of geometric quantities, e.g. the 4-volume of each
4-simplex, reduces to a counting problem; we will analyse in details
the causal set reformulation of the model in the following section.

\item Also, if we fix all the geometric data to be those obtained from
a fixed edge length for any edge in the triangulation (e.g. the Planck
length), then what we obtain is the conventional and well-studied sum
over triangulations in the dynamical triangulations approach to
quantum gravity, for causally well-behaved Lorentzian triangulations; the
additional integrals, if instead left to be performed, can be
intepreted as providing a sum over proper times that is usually not
implemented in that approach; this may leave hopes for the expected
emergence of a smooth classical limit from the present simplicial
model, in light of the important results obtained recently
\cite{amblol1}\cite{amblol2}\cite{amblol3} on Lorentzian dynamical triangulations.

\end{itemize}

Let us finally comment on the analogy with the relativistic particle
case. We have seen that different quantum amplitudes may be
constructed for a relativistic particle: the Wightman function,
solution of all the constraints (defining a projector operator), distinguishing the order of its
arguments, and taking into account all the trajectories of interest
for a single paticle (not anti-particle); the Hadamard function,
solution of all the constraints, not registering the order of its
arguments, and including all the trajectories of both particle and
anti-particle solutions; the Feynman propagator, not a solution of the
constraint, registering the order of its arguments, again
including both particle and anti-particle solutions, and which reduces
to the appropriate Wightman function depending on the time-ordering of
its arguments. It is clear that for a single-particle theory and if we
do not consider anti-particles, then the Wightman function is all we
have, and all we need to define a physical transition amplitude
between states.  The Feynman propagator is of not much use in a theory of a
single particle, but needed in a theory of many particles. If we want
instead something which is a
solution
to the constraint, but not reflect any ordering, then the only answer is the Hadamard
function. It can be interpreted as a particle plus an
antiparticle amplitude, but the point is really just the ordering of the
arguments in the amplitude. Now, the question we have to ask in our
spin foam case is: is it a single particle or a multiparticle theory? does the amplitude we have in the
Barrett-Crane model reflect the ordering of the arguments or not? if we
impose it to reflect such an ordering, what kind of amplitude do we get?

Now, the BC amplitude is real and does not reflect any ordering, so it
cannot correspond to a Wightman function, and it must instead correspond
to the Hadamard function (for what the particle analogy is useful). The
one may say that in order to impose causality (ordering-dependence), I have
to reduce it to the Wightman function. In any case this will give us
something different from the Barrett-Crane amplitude as it is. How to
impose the restriction, then? The only possible
restriction imposing an ordering dependence seem to be the one we performed. This is basically
because in the spin foam we have only combinatorial data and
representations, and the only ordering we can impose is an ordering at the
level of the 2-complex. If we do it, and ask for an amplitude that
reflects it, we get the above causal amplitude.

Now, is there a way to interpret what we do in terms of the particle
analogy? Work on this issue is in progress, and the idea would be that what we have is a
multiparticle theory, where we have a particle on the
hyperboloid corresponding to each simple representation (there is indeed
an interpretation of the kernel $K$ as a green function for a particle on
the hyperboloid); then the two parts of the kernel would
correspond to the two Wightman functions and the kernel itself to the Hadamard
function; so the two different orientations for the same
particle-face-representation correspond to something like a particle and
an antiparticle, or to the two Wightman functions $G^+$ and $G^-$. In this context the right amplitude
would be exactly that selecting a particle for one orientation and an
antiparticle for the opposite orientation, i.e. simply one or the
other Wightman function. This is what we do, in fact, and this is what the
Feynman propagator does in the particle case.

Of course we should bear in mind that we should not rely too much on perturbative QFT, nor on the
particle analogy, but nevertheless they can give some insight;
moreover, we know from the group field theory that in some non-trivial
sense what we are dealing with is indeed a form of perturbative field
theory, and that the spin foam
for fixed 2-complex is exactly a perturbative
Feynman graph, although very peculiar.

\subsection{Barrett-Crane model as a quantum causal set model}
In the new formulation presented above, the Barrett-Crane model fits
into the general scheme of quantum causal sets (or quantum causal
histories) models, as developed in \cite{fotlee}\cite{fot}\cite{fot2}\cite{set}. It
represents, in fact, the first explicit and highly non-trivial example
of such a class of models. As
previously said,
all the metric information of a Lorentzian space-time can
be encoded in its causal set skeleton apart from its conformal factor,
and therefore causal sets are a natural context for a discrete and
finitary
space-time model \cite{Sorkin}\cite{Sork}. Reformulating the
Barrett-Crane model in these terms helps in understanding the
underlying causal structure in the model, and gives insight on the
properties required for quantum causal
sets (especially the evolution operators).

\subsubsection{General framework for quantum causal sets}

Let us first recall the basic elements of this scheme.

Consider a discrete set of events $\{p, q, r, s, ...\}$, endowed
with an ordering relation $\leq$, where the equal sign implies
that the two events coincide. The ordering relation is assumed to
be reflexive ($\forall q, q\leq q$), antisymmetric ($q\leq s,
s\leq q \Rightarrow q=s$) and transitive ($ q\leq r, r\leq s\,
\Rightarrow q\leq s$). These properties characterize the set as a
partially ordered set, or poset, or, interpreting the ordering
relation as a causal relation between two events, as a causal set
(see Fig.4). In this last case, the antisymmetry of the ordering
relation, together with transitivity, implies the absence of
closed timelike loops. We also report a few definitions that will
be useful later:
\begin{itemize}
\item the causal past of an event $p$ is the set $P(p)=\{ r | r\leq p\}$;
\item the causal future of an event $p$ is the set $F(p)=\{ r | p\leq r\}$;
\item an acausal set is a set of events unrelated to each other;
\item an acausal set $\alpha$ is a complete past for the event $p$ if
$\forall r\in P(p), \exists s\in\alpha \,|\, r\leq s\,\,or\,\,s\leq r$;
\item an acausal set $\beta$ is a complete future for the event $p$ if
$\forall r\in F(p), \exists s\in\beta \, |\, r\leq s\,\,or\,\,s\leq r$;
\item the causal past of an acausal set $\alpha$ is
$P(\alpha)=\cup_i\,P(p_i)$ for all $p_i\in\alpha$, while its causal
future is $F(\alpha)=\cup_i\,F(p_i)$ for all $p_i\in\alpha$;
\item an acausal set $\alpha$ is a complete past (future) of the
acausal set $\beta$ if $\forall p\in P(\beta) (F(\beta))\,\,\exists
q\in\alpha \,\,|\,\,p\leq q\,\,or\,\,q\leq p$;
\item two acausal sets $\alpha$ and $\beta$ are a complete pair when
$\alpha$ is a complete past of $\beta$ and $\beta$ is a complete future
for $\alpha$.
\end{itemize}

From a given causal set, one may construct another causal set
given by the set of a-causal sets within it endowed with the
ordering relation $\rightarrow$, so that $\alpha\rightarrow\beta$
if $\alpha$ and $\beta$ form a complete pair, also required to be
reflexive, antisymmetric and transitive. This poset of a-causal
sets is actually the basis of the quantum histories model.

The quantization of the causal set is as follows (see Fig.4).
It can be seen as a functor
between the causal set and the categories of Hilbert spaces.

\begin{figure}
\begin{center}
\includegraphics[width=9cm]{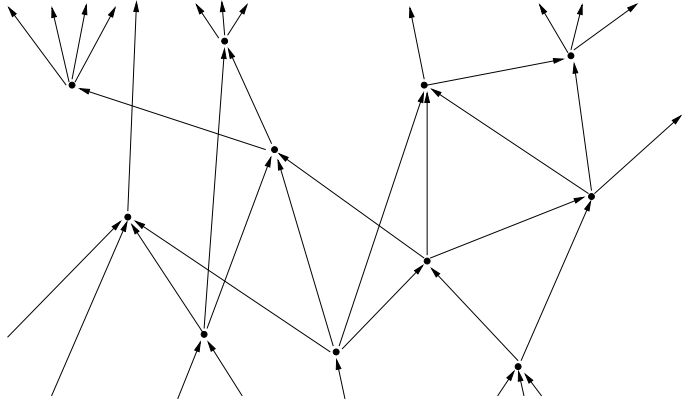}
\caption{A causal set}
\end{center}
\end{figure}

We attach an Hilbert space $\mathcal{H}$ to each node-event and
tensor together the Hilbert spaces of the events which are
spacelike separated, i.e. causally unrelated to each other; in
particular this gives for a given a-causal set
$\alpha=\{p_i,...,p_i,...\}$ the Hilbert space
$\mathcal{H}_\alpha=\otimes_i \mathcal{H}(p_i)$.

Then given two a-causal sets $\alpha$ and $\beta$ such that
$\alpha\rightarrow\beta$, we assign an evolution operator  between
their Hilbert spaces: \be
E_{\alpha\beta}\,:\,\mathcal{H}_{\alpha}\,\rightarrow\,\mathcal{H}_{\beta}.
\ee

In the original Markopoulou' scheme, the Hilbert spaces considered
are always of the same (finite) dimension, and the evolution
operator is supposed to be unitary, and fully reflecting the
properties of the underlying causal set, i.e. being reflexive:
$E_{\alpha\alpha}=Id_\alpha$, antisymmetric:
$E_{\alpha\beta}E_{\beta\alpha}=Id_\alpha
\Leftrightarrow\,E_{\alpha\beta}=E_{\beta\alpha}=Id_\alpha$, and
transitive: $E_{\alpha\beta}E_{\beta\gamma}=E_{\alpha\gamma}$.

The evolution operators mapping complete pairs that are causally
unrelated to each other, and can be thus tensored together, may also
be tensored together, so that there are cases when given evolution
operators may be decomposed into elementary components. More on the
dynamics defined by this type of models can be found in \cite{fot}\cite{fot2}.

Another possibility is to assign Hilbert spaces to the causal
relations (arrows) and evolution operators to the nodes-events.
This matches the intuition that an event in a causal set (in
spacetime) is an elementary change. This second possibility gives
rise to an evolution that respects local causality, while the
first one does not. This is also the possibility realised in the
causal Barrett-Crane model, as we are going to discuss. Also, this
hints at a fully relational reformulation of quantum mechanics
\cite{Carlo}\cite{Smolin}, since the Hilbert space between two
events, $a$ and $b$, admits the interpretation of describing the
possible states of ``$a$ seen by $b$'' (and reciprocally); this is
closely related also to the ``many-views'' category-theoretic
formulations of sum-over-histories quantum mechanics \cite{Isham}.

\subsubsection{The quantum BC causal set}

Let us consider now how this scheme is implemented in the causal
Barrett-Crane model defined above.

The starting point is an oriented graph, i.e. a set of vertices linked
by arrows, restricted to be 5-valent (each vertex is source or target
of five arrows) and to not include closed cycles of
arrows. Interpreting the arrows as representing causal relations, this
is a causal set as defined above, the orientation of the links
reflects the ordering relation among the vertices, and it does not contains closed
timelike curves.

Because of the restriction on the valence, it can be decomposed into building blocks, in the sense that for each
vertex, only one out of four possibilities may be realised, depending
on how the arrows are related to the vertex itself (see Fig.5).

\begin{figure}
\begin{center}
\includegraphics[width=9cm]{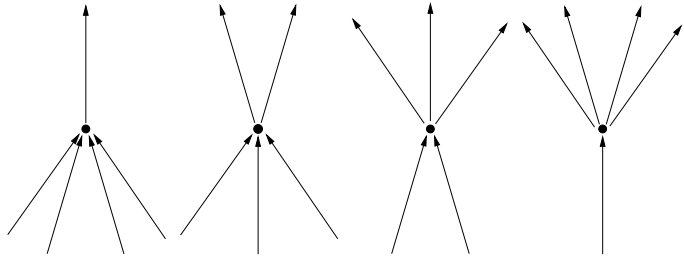}
\caption{Building blocks for the BC causal set}
\end{center}
\end{figure}

This causal set is just the first {\it stratum} of the 2-complex on
which the Barrett-Crane Lorentzian spin foam model is based, assuming it
to be oriented. As such, there is a dual simplicial interpretation of
the elements in it, since now the vertices are dual to 4-simplices and
the oriented links are dual to oriented tetrahedra, the tetrahedra in
a given 4-simplex can be considered as belonging to different
spacelike hypersurfaces so that the four
moves (in Fig.5, above) represent the four possible Pachner moves ($4-1$, $3-2$, and
their reciprocal) giving the evolution of a 3-dimensional simplicial
manifold in time.

The crucial point, we note, is the identification of the direction of
the arrow in the causal set with the orientation of the tetrahedron it
refers to.

The quantization is then the assignment of Hilbert spaces, the
Hilbert spaces of intertwiners among four given continuous
(principal irreducible unitary) representations of the Lorentz
group, representing the possible states of the tetrahedra in the
manifold, to the {\it arrows} of the causal set (oriented edges of
the 2-complex)and of the causal Barrett-Crane amplitude defined
above to the nodes of the causal set, as the appropriate evolution
operator between the Hilbert spaces. We have seen in fact how it
registers the ordering of its arguments and how it can thus be
interpreted as the right transition amplitude between spin network
states.

More precisely it proceeds as follows.

We pass from the causal set defined above to the so-called {\it
edge-poset}, i.e. the causal set obtained associating a node to
each arrow of the previous one and a new causal relation (arrow)
linking each pair of nodes corresponding to not spacelike
separated arrows in the previous graph. The building blocks of
this new causal set, obtained directly form those of the previous
poset, are then as in Fig.6.

\begin{figure}
\begin{center}
\includegraphics[width=9cm]{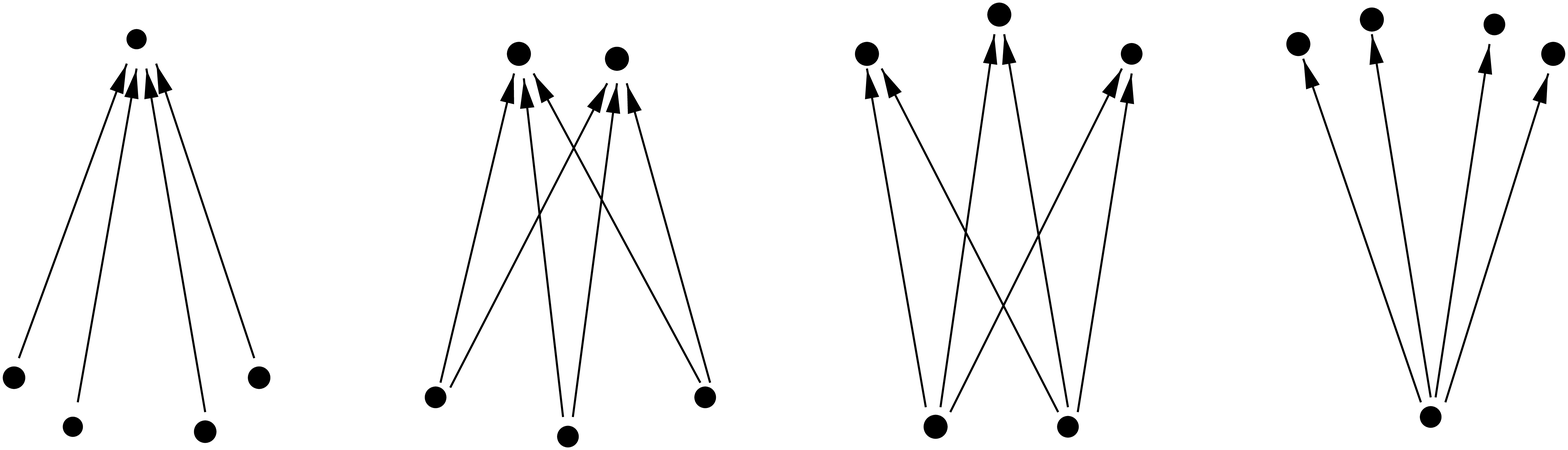}
\caption{Building blocks for the BC edge-poset}
\end{center}
\end{figure}
To each node in this new causal set it is associated the Hilbert
space of intertwiners between four simple representations of the
Lorentz group. More precisely, an ``intertwiner'' here is the open (simple)
spin network with a single vertex and with a fixed normal $x$ as described
in equation \Ref{xintertwiner},
so that it is labelled by four vectors living in the four representations
and the normal $x\in{\cal H_+}$ sitting at the vertex.
The Hilbert space of such ``interwiners'' is then
$L^2({\cal H}_+)_{\rho_1,\rho_2,\rho_3,\rho_4}$, where
we have specified the four intertwined representations. 
As we said above, Hilbert spaces that are
a-causal to each other can be tensored together. In particular, in
each of the building blocks the source nodes form a minimal
a-causal set and the target nodes as well; moreover, these form a
complete pair, so they are in turn linked by a causal relation in
the poset of a-causal sets defined on the edge-poset; to each of
these causal relations among complete pairs, i.e. to each of the
building blocks of the edge-poset, we associate the causal
Barrett-Crane amplitude defined earlier \Ref{orientampli}.

Taking two touching (but a-causally related) tetrahedra or
correspondingly two interwtiners, we can tensor them and glue them together.
Doing so, we get an open (simple) spin network with two vertices describing
the state of the system formed by these two tetrahedra. More precisely, let's note
$x_1$ ($x_2$) the normal of the first (second) intertwiner,
$\alpha_1,\dots,\alpha_4$ ($\beta_i$) its four representations and
$(j_i,m_i)$ ($(k_i,n_i)$) the vectors in the $\alpha_i$ ($\beta_i$)
representations living at the ends of the one-vertex spin network.
Let's suppose that the two intertwiners are glued along the edge
$\rho=\alpha_1=\beta_1$, then we need to identify the vectors
$(j_1,m_1)=(k_1,n_1)$ and sum over them. This yields the right
open spin networks with two vertices. At the level of the normals,
the resulting tensored Hilbert space
is $L^2(x_1\in{\cal H}_+)\times L^2(x_2\in{\cal H}_+)$.
In particular, the state
corresponding to any of the a-causal sets present in the building
blocks is given by an open simple spin network, with
representations of $SO(3,1)$, but also representations of a given
$SU(2)$ subgroup and a label (``angular momentum projection'') for
states in this subgroup on their open ends, and the causal
Barrett-Crane amplitude for a single vertex (4-simplex), with
appropriate boundary terms, interpolates between these states.

Of course, just as in the general case, states (and operators
between them) referring to a-causal sets that are not causally
related to each other can be tensored together again, and
composite states constructed in this way evolve according to
composite evolution operators built up from the fundamental ones.

Let us now discuss what are the properties of the evolution operator
(Barrett-Crane causal amplitude) in this quantum causal set context.

First of all we recall that in the original Markopoulou' scheme
\cite{fot} the single poset evolution operator $E_{\alpha\beta}$
is required to be reflexive, antisymmetric, transitive and unitary
(we have defined above what these properties mean in formulae).

However, we have stressed from the beginning that the
Barrett-Crane partition function for a given 2-complex (triangulation)
has to be understood as just one term in a sum over 2-complexes of all
the partition functions associated with them (with some given
weight). In this causal set picture this means that we have to
construct a sum over causal sets interpolating between given boundary
states, as outlined also in \cite{fot} and each of them will be then
weighted by the causal Barrett-Crane amplitude, made out of the
building block evolution operators as explained.

Therefore, given two a-causal sets $\alpha$ and $\beta$, the full
evolution operator between them will be given by an operator
$\mathcal{E}_{\alpha\beta}$ defined as: \be
\mathcal{E}_{\alpha\beta}\,=\,\sum_c\,\lambda_c\,E_{\alpha\beta}^c
\ee where we have labelled the previously defined evolution
operator for a single causal set with $c$ to stress its dependence
on the underlying graph, and we also included a possible
additional weight $\lambda_c$ for each causal set in the sum.
Recall also the formal expression for a path integral for quantum
gravity we outlined in the introduction, where we split the sum
over geometries into a sum over causal structures (causal sets)
and a sum over metric degrees of freedom defining a length scale
in a consistent way.

It is clear then that this is the real physical evolution operator
between states, and this is the operator whose physical properties
have to be understood\footnotemark \footnotetext{Let us note that
there is another possible way of getting rid of the dependence of
the theory on any fixed underlying causal set, which is to define
the complete model by a refining procedure of the initial causal
set, that increases progressively (possibly to infinity) the
number of vertices in it and thus defines the model as the limit
of the fixed poset one under this refinement. We tend to prefer
the solution provided by the sum over 2-complexes (causal sets)
only because, while there exist (acausal) models that furnish this
last sum, i.e. those based on the formalism of group field theory,
there is (to the best of our knowledge) no implementation yet for
this refining procedure. Also the first solution seems to us more
in agreement with the general idea of summing over geometries to
obtain the amplitudes for a quantum gravity theory}.

It is sensible to require the partial operators $E_{\alpha\beta}$ to
be reflexive, of course, since imposing this leads to an analogue
property for the full evolution operator (up to a factor dependng on
$\lambda$).

Also, we require both the fixed-poset evolution operator $E^c$ and the
full evolution operator $\mathcal{E}$ to be antisymmetric, to reflect
the property of the underlying causal set, in the first case, and as a
way to implement the physical requirement that imposing that the
evolution passes through a given state changes significantly the
evolution of a system, even if it then returns in the initial state.

For the other properties the situation is more delicate.

Doubts on the meaningfulness to require transitivity for the
single poset evolution operator were put forward already in
\cite{fot}. The reason for this is that this properties implies a
sort of directed triangulation invariance of the resulting model,
in that the evolution results in being at least partly independent
on the details of the structure of the causal set itself. In fact,
transitivity implies that, as far as the evolution operators are
concerned, a sequence of two causal relations linking the a-causal
sets $\alpha$ and $\beta$ and then $\beta$ and $\gamma$ is
perfectly equivalent to a single arrow linking directly $\alpha$
and $\gamma$. The causal set is in turn dual to a triangulation of
the manifold and this is why transitivity implies a sort of
triangulation invariance.

Therefore, while it is certainly meaningful to require transitivity
even for the single-poset evolution operators $E^c_{\alpha\beta}$ if
we are dealing with 3-dimensional gravity, which is a topological
field theory, it does not seem appropriate to impose it also in the
4-dimensional case, where any choice of a fixed causal set is a
restriction of physical degrees of freedom of the gravitational field.

On the other hand, the situation for the full evolution operator is
different. Even in the non-topological 4-dimensional case, we do
require that the evolution from a given acausal set $\alpha$ to
another $\gamma$ is independent of the possible intermediate states in
the transition, {\it if we sum over these possible intermediate
states}, i.e. we require:

\be
\sum_\beta\mathcal{E}_{\alpha\beta}\mathcal{E}_{\beta\gamma}\,=\,\mathcal{E}_{\alpha\gamma},
\ee which is nothing else than the usual composition of amplitudes
in ordinary quantum mechanics. We do not require the same property
to hold if we drop the sum over intermediate states (an
intermediate state corresponding to an intermediate  measurement
of some physical properties).

The property of unitarity of the evolution is of extreme
importance. We believe that the evolution defined by the causal
spin foam model has to be indeed unitary to be physically
acceptable.

In fact, the initial and final a-causal sets that represent the
arguments of the evolution operator, i.e. of the causal amplitudes
of the model, are always assumed to form a complete pair, and this
completeness should hold for any causal set summed over in the
definition of the full evolution operator mapping one a-causal set
into the other. Therefore, if the final a-causal set is a complete
future of the first, and the initial a-causal set is a complete
past for the final one (this is, we recall, the definition of a
complete pair), we expect all the information from the initial
state to flow to the final state. The requirement of unitarity is
thus nothing more than the requirement of the conservation of
information in the evolution.

We stress again, however, that the physical evolution operator is
$\mathcal{E}$, and not $E^c$, i.e. it involves a sum over causal sets.
Any term in this sum, on the contrary, represents a truncation of the
full dynamical degrees of freedom in the theory, and a restriction on
the flow of information.
Therefore, it is this operator that, we think, has to be required to be unitary.

But if this is the case, then the single-poset operator $E$ has to be
{\it not} unitary, as it is easy to verify (a sum of unitary operators
cannot be a unitary operator itself).

Therefore we require:
\be
\sum_{\beta}\mathcal{E}_{\alpha\beta}\,\mathcal{E}^{\dagger}_{\alpha\beta}\,=\,
\sum_\beta\mathcal{E}_{\alpha\beta}\bar{\mathcal{E}}_{\beta\alpha}\,=\,Id_\alpha
\ee

which implies:
\be
\sum_\beta E^c_{\alpha\beta}\,E^{c\dagger}_{\alpha\beta}\,=\,
\sum_\beta E^c_{\alpha\beta}\bar{E}^c_{\beta\alpha}\,\neq\,Id_\alpha.
\ee

\begin{figure}
\begin{center}
\includegraphics[width=6cm]{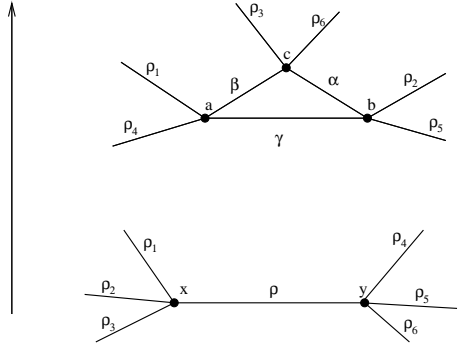}
\caption{4-simplex: the $2\rightarrow 3$ move}
\end{center}
\end{figure}

More precisely, let's consider a $2\rightarrow 3$
move as on Fig.7 and its amplitude read directly from the oriented
spin foam amplitude \Ref{orientampli} (choosing $\mu=+1$ for the
global orientation of the 4-simplex):
\bes
{\cal A}(2\rightarrow 3)
&=&E^c_{\rho\rightarrow(\alpha,\beta,\gamma)} \nonumber \\
&=&
\int \textrm{d}x\,\textrm{d}y\,\textrm{d}a\,
\textrm{d}b\,\textrm{d}c \quad
\prod_{v,w\in\{x,y,a,b,c\}}\sinh^{-1}(\theta(v,w)) \nonumber \\
&&e^{+i\rho\theta(x,y)}
e^{+i\alpha\theta(b,c)}
e^{+i\beta\theta(c,a)}
e^{+i\gamma\theta(a,b)} \times \nonumber \\
&&e^{-i\rho_1\theta(x,a)}
e^{-i\rho_2\theta(x,b)}
e^{-i\rho_3\theta(x,c)}
e^{-i\rho_4\theta(y,a)}
e^{-i\rho_5\theta(y,b)}
e^{-i\rho_6\theta(y,c)}.
\ees
In fact, when you ``acts'' with this (oriented) 4-simplex on the region of a
(boundary) simple spin network made of two glued tetrahedra, one has to
take into account the shift between the two normals to the tetrahedra on the boundary
state and their values on the 4-simplex, so that the final evolution operator consists
of, first, Lorentz transformations to go to the initial normals to the normals attached
to the 4-simplex and then the transformation from a 2 vertex open spin network to a
3 vertex open spin network generated by the 4-simplex itself. This way, we recover the full
oriented spin foam amplitude with the eye diagram weights.
Changing the time orientation, we simply change $\mu$ from $1$ to $-1$, which
changes the above operator to its complex conjugate.

Let us now finally check that the causal amplitudes defined above
have all the properties we want (or not want) them to have.

We do not have at our disposal a well-motivated definition of the
complete evolution operator, coming from well understood physical
requirements or obtained within some known mathematical formalism,
e.g. a group field theory model. Therefore we must limit ourselves to
check the wanted (and not-wanted) properties of the fixed-poset
evolution operators. We will indeed check these properties in the
simplest case of the evolution operator associated to a single
building block of the causal set, i.e. a single choice of minimal
source and target acausal sets.

We recall that the properties we want such an operator to satisfy are:
reflexivity, antisymmetry, absence of transitivity and absence of
unitarity.
We satisfy reflexivity trivially just by defining the evolution
operator $E_{\alpha\alpha}$ to be the identity operator.

As for antisymmetry, we clearly have $E_{\alpha\beta}E_{\beta\alpha}\ne \textrm{Id}$.

Transitivity is NOT satisfied. Indeed, composing two  4-simplex
operators is not equivalent to a single 4-simplex operator. This is
straightforwardly due to the non-isomorphism of the initial and final (boundary)
states of a 4-simplex move.

Finally, most relevant, unitarity is NOT satisfied either.
Indeed starting from two open spin networks, made of 2 vertices or equivalently
2 glued tetrahedra, then the resulting states after a $2\rightarrow 3$ move should have
a identical scalar to the one of the initial states. Considering that the scalar product
of the initial spin networks is $\delta(\rho-\rho')$ and that the scalar product
of two 3 vertex open spin network is
$\delta(\alpha-\alpha')\delta(\beta-\beta')\delta(\gamma-\gamma')$, it is
straightforward to check that this is not the case.

The unitarity is reserved to the
evolution operator resulting from a sum over intermediate traingulations. Nevertheless,
unitarity is equivalent to conservation of the information. Intuitively, this is
violated by the fact that the Hilbert space of 2 vertex open spin networks is NOT
isomorphic to the one of 3 vertex open spin networks: they do not carry the same
information (not same number of internal representations). Is there still a way in which
unitarity is verified at this microscopic level. The answer resides in the
edge poset picture. Indeed, looking at Fig.6, each arrow represents the same 4-simplex, but
each carry a different amplitude/operator which is a single exponential. This operator
takes the normal of the past tetrahedron of the arrow to the normal of the future
tetrahedron, and thus is an automorphism of $L^2({\cal H}_+)$.
This way, we can define an unitary evolution attached to the 4-simplex
but it
means associating not one but many unitary operators to it.

This concludes the formulation of the causal model based on the
causally restricted Barrett-Crane amplitudes as a quantum causal set model.

\section{Conclusions and outlooks}
Let us summarize what are the results of this work.

We have studied the geometrical meaning of the variables that are
present in the Barrett-Crane model and, consequently, what kind of
classical description of spacetime geometry is hidden in its
formulation and what action should describe the classical limit of
the model; this led also to a possible explanation of the reason
why the model associates two distinct normals to each tetrahedron
in the interior of the manifold. We have analysed the quantum
geometry of the model, identifying the Hilbert space of boundary
states and its properties, and discussing the meaning of the
amplitudes appearing in the partition function. We explained why
the Barrett-Crane model has to be considered as providing the
physical inner product between quantum gravity states, i.e. as
realizing in a covariant manner (as a sum-over-histories) the
projector operator onto physical states, and we identified
explicitly in the model the $Z_2$ symmetry that characterizes such
a covariant implementation of the projector.

Moreover, we have shown how it is possible to break in a
consistent way this symmetry and obtain a spin foam realization of
the quantum gravity analogue of the Feynman propagator, i.e. a
causal transition amplitude between quantum gravity states
represented by simple spin networks. The resulting spin foam model
turns out to be a path integral for Lorentzian first order Regge
calculus where all the geometric variables are expressed in a
purely algebraic way and with a clear definition of the measure
for the integration variables. It is the first explicit example of
a causal spin foam model. In this way, the resulting causal model
fits into the general framework of quantum causal sets (or quantum
causal histories), as we described in detail, and represents the
first (highly) non-trivial example of these in 4 dimensions.

The causal model we propose is thus a link among several areas of
research: topological field theory, canonical quantum gravity,
sum-over-histories formulation of quantum gravity, Regge calculus,
causal sets and dynamical triangulations.

There are many outlooks to the present work.

As we have shown, there is a clear link between causality in spin
foam models and the orientation of the 2-complex on which they are
based. The causal restriction we performed on the Barrett-Crane
model was indeed motivated by such a link. The resulting causal
model, however, was, on the one hand, constructed in an admittedly
rather ad hoc manner and not derived from some general formalism,
and, on the other hand, was restricted to a choice of a fixed
triangulation of the spacetime manifold, i.e. to a fixed
2-complex. The present derivations of the Barrett-Crane model,
avoiding this limitation, are all based on the formalism of group
field theory and define a sum over 2-complexes in a very natural
way. However, they all generate un-oriented 2-complexes and thus
the a-causal Barrett-Crane model. One could try to obtain the
previous causal model from a group field theory and we believe
that the best solution would be to use a complex field over
$\sl^4$ or else to use complex representations of $\sl$,
thus generalising the existing formalism. This possibility is
currently being investigated.

It would also be interesting to study more in detail the causal
$2+1$ spin foam model based on the discrete (positive)
representations of $SU(1,1)$ and study its precise relation with
$2+1$ general relativity.

As we pointed out, the causal model constructed in this paper can
be seen as a quantum causal set model, but also as a
generalization of the dynamical triangulation approach, in that it
contains more degrees of freedom than these, allowing dynamical
``length scale'' data. It is thus possible to expect a
cross-fertilization of these areas of quantum research, with
results and techniques from causal set theory and from the
dynamical triangulation approach to be imported in the causal spin
foam framework. This may help in solving the many problems it
still faces, such as the definition of a semi-classical
approximation or of a continuum limit, or a better understanding
of the structures involved in it, to name but a few.

From the point of view of quantum causal histories, many issues
are to be investigated and the model presented may be where to
analyse these issues in concrete and explicit terms; in
particular, much has to be understood about the role of
transitivity and unitarity, about the definition of observables
and about the application of the general framework to problems in
quantum cosmology. Moreover, the relational point of view on
quantum mechanics, that these models support, has to be developed
and investigated in much more details.

At the classical level, the first order version of Regge calculus
advocated here as the classical theory of simplicial gravity
corresponding to spin foam models has also to be studied in depth.

Another direction for further research is to look at a quantum
deformed version of the Barrett-Crane model. Indeed it seems that
it amounts to discretizing the normal $x$ to the tetrahedra, which
would then live in a quantum hyperboloid made from piling fuzzy
spheres \cite{roche}. This quantum model would be relevant to the
present discussion since it is supposed to be related to the
constrained $BF$ theory with a cosmological constant \cite{q}:

\be
S_\Lambda[B,A]= \int B \w F +\phi(\textrm{constraints}) +\Lambda B\w B
\ee
Contrarily to the 3d case, where the action with cosmological constant
changes sign under $B\rightarrow -B$ \cite{3dvolume},
it is easy to see that in this case:

\be S_\Lambda[-B]=-S_{-\Lambda}[B]\ne-S_\Lambda[B] \label{Slambda}
\ee So it could be useful to distinguish these two sectors of the
theory and the link with causality issues in the presence of the
cosmological constant, since the situation seems more involved
than in the case studied here. A first step would be to check the
relation between the $q$ deformation parameter and the
cosmological constant $\Lambda$ through, most likely, an
asymptotic expansion of the quantum $\{10\rho\}$ symbol.

Further along the line, along the plan outlined in the
introduction, is the study of the statistical mechanics of the
degrees of freedom of the causal spin foam model, and of the
consequent thermodynamics; in particular, one should apply the
model to the study of black hole thermodynamics, having first
obtained a purely algebraic and combinatorial characterization of
causal horizons, and try to obtain the known relation between
entropy and area of a black hole, in the appropriate
semi-classical regime. It is reasonable to expect that the
holographic principle \cite{Bousso}\cite{MarSmo}, in a backgound
independent and algebraic formulation to be found, will play a
major role here. If this turns out to be the possible, then much
more confidence can be put on the belief that the model reproduces
the dynamics of spacetime geometry given by general relativity,
and that we are on the right track.

\section*{Acknowledgements}
We would like to thank R. M. Williams, C. Rovelli and H. Pfeiffer
for discussions and comments on the manuscript. 
D.O. acknowledges financial support from EPSRC, Cambridge European Trust and Girton College. 


\end{document}